\documentclass[11pt,a4paper]{article} 
\pdfoutput=1
\usepackage{jcappub}
\usepackage{bm}
\usepackage{times} 
\usepackage[utf8]{inputenc}
\usepackage[english]{babel}
\usepackage[T1]{fontenc}
\usepackage{graphicx}
\usepackage{amsmath}
\usepackage{subfig}
\usepackage{hyperref}
\usepackage{color}
\usepackage{amssymb}
\usepackage[font=scriptsize,labelfont=bf]{caption}
\usepackage{appendix}
\usepackage{slashed}
\usepackage{slantsc}
\usepackage{braket}
\usepackage{mathrsfs}
\usepackage{subfig}
\usepackage{placeins}
\usepackage{fullpage}
\usepackage{sidecap}
\usepackage{float}
\usepackage{wrapfig}
\usepackage[dvipsnames]{xcolor}
\usepackage{dsfont}
\usepackage{framed}
\usepackage{soul}
\usepackage{comment}
	
\begin{document}

\begin{titlepage}

\begin{center}

\vskip 1.0 cm

{\LARGE \bf Galaxy Two-Point Correlation Function in General Relativity}

\vskip 1.0 cm

{\large
Fulvio Scaccabarozzi$^{a}$, \,Jaiyul Yoo$^{a,b}$ and Sang Gyu Biern$^{c,a}$	
}

\vskip 0.5cm

{\it
$^{a}$Center for Theoretical Astrophysics and Cosmology,
Institute for Computational Science, University of Z\"urich, Winterthurerstrasse 190, CH-8057, Z\"urich, Switzerland
\vspace{0.2 cm}
\\
$^{b}$Physics Institute, University of Z\"urich, Winterthurerstrasse 190, CH-8057, Z\"urich, Switzerland
\vspace{0.2 cm}
\\
$^{c}$Optotune, Bernstrasse 388, CH-8953, Dietikon, Switzerland
}

\vspace{1 cm}

\today

\vskip 1.0 cm

\end{center}

\vspace{1.5cm}
\hrule \vspace{0.3cm}
\noindent {\sffamily \bfseries Abstract} 
\\
\\
We perform theoretical and numerical studies of the full relativistic
two-point galaxy correlation function, considering the linear-order
scalar and tensor perturbation contributions and the wide-angle effects.
Using the gauge-invariant relativistic
description of galaxy clustering and accounting
for the contributions at the observer position, we demonstrate that the
complete
theoretical expression is devoid of any long-mode contributions from scalar
or tensor perturbations and it lacks the infrared divergences in agreement
with the equivalence principle. By showing that the gravitational potential contribution to the correlation function converges in the infrared, our study justifies an IR cut-off $(k_{\text{IR}} \leq H_0)$ in computing the gravitational potential contribution. Using the full gauge-invariant expression,
we numerically compute the galaxy two-point correlation function
and study the individual contributions in the conformal Newtonian gauge.
We find that the terms at the observer position such as the coordinate lapses and the observer velocity (missing in the standard formalism) dominate over the other relativistic contributions  in the conformal Newtonian gauge such as the source velocity, the gravitational potential, the integrated Sachs-Wolf effect, the Shapiro time-delay and the lensing convergence. Compared to the standard Newtonian theoretical predictions
that consider only the density fluctuation and redshift-space distortions, the relativistic effects in galaxy clustering result in a few
percent-level systematic errors beyond the scale of the baryonic acoustic
oscillation ($\sim 2\%$ at 150 Mpc/h and redshift one). Our theoretical and numerical study
provides a comprehensive understanding of the relativistic effects in the
galaxy two-point correlation function, as it proves the validity of the theoretical prediction and accounts for effects that are often neglected in its numerical evaluation.

\vskip 10pt
\hrule
\vspace{0.6cm}
\end{titlepage}

\pagebreak

\noindent\hrulefill
{\linespread{0.75}\tableofcontents}
\noindent\hrulefill

\section{Introduction}

Galaxy surveys map the universe by measuring the redshift $z$ and the direction $\boldsymbol{\hat n}$ of each galaxy.  
One simple and direct way to extract physical information from this map is then to compute the galaxy two-point correlation function.
In particular, one correlates the number density of galaxies in a redshift bin around $z_1$ and in a small solid angle around a direction $\boldsymbol{\hat n}_1$ with those in a redshift bin around $z_2$ and in a small solid angle around a direction $\boldsymbol{\hat n}_2$. 
The next generation of galaxy surveys \cite{Laureijs:2011gra,Amendola:2016saw,Abate:2012za,Abell:2009aa,Aghamousa:2016zmz,Carilli:2004nx} will probe the large scale structure of the universe at high redshift and for wide regions of the sky.
Given the unprecedented precision achieved by the recent observational advances, the theoretical predictions of the two-point correlations used to analyze the data can no longer rely on the flat-sky approximation nor on the assumption that the universe is homogeneous and isotropic.
The flat-sky approximation, assuming that the directions $\boldsymbol{\hat n}_1$ and $\boldsymbol{\hat n}_2$ coincide, is currently used to analyze redshift surveys and constrain cosmological parameters but is not sufficiently accurate to interpret data from future surveys \cite{Okumura:2007br}. Furthermore, most expressions used for the analysis only take into account density fluctuations and redshift-space distortions. Clearly, these standard expressions provide an approximation to what we observe, and they are inevitably gauge-dependent. Indeed, a gauge-invariant expression of the two-point correlation function includes all relativistic effects that manifest in galaxy clustering.

Previous works have studied the impact of relativistic effects on the correlation function and the power spectrum (as well as additional subdominant effects such as \cite{Tansella:2018tbo,Baumann:2017lmt}). 
In \cite{Yoo:2012,Jeong,Yoo:2010}, the galaxy power spectrum was derived,
including all the relativistic effects, and its
detection significance was quantified. However, they adopted the flat-sky
approximation, essentially ignoring the relativistic effects
at the observer position and along the line-of-sight direction, when
computing the detection significance by using the power spectrum. Since the Fourier decomposition
is non-local in nature, the standard power spectrum has difficulty in its
expression in the all-sky limit (see, however, \cite{YODE13,Heavens:1994iq,Binney,Lahav:1994xp} for the all-sky
analysis using the spherical Fourier decomposition). However, the correlation
function is subject to no such complications and can be derived directly
in terms of observable quantities. Using the standard redshift-space
distortion formula, the galaxy two-point correlation function was derived
in \cite{Szalay:1997cc,Szapudi:2004gh,Papai:2008bd,Raccanelli:2010hk,Samushia:2011cs,Slepian:2015tna,Matsubara1,Matsubara2} without assuming the flat-sky approximation.
In light of the full relativistic description of galaxy clustering 
\cite{Yoo:2009,Yoo:2010,Yoo}, a complete description of the galaxy
two-point correlation function was derived \cite{Yoo:2013}, while ignoring
the gravitational potential contributions, but finding several new
corrections from the velocity perturbations. 

In recent years, many efforts have been made to
compute the galaxy two-point correlation function with all the relativistic
effects (see, e.g., \cite{Bertacca:2012tp,Raccanelli:2013dza, 
Raccanelli:2013gja,Raccanelli:2015vla}).  For example,
the lensing effect  arises from the
matter density fluctuations along the line-of-sight direction, and its
contribution to the correlation function has been studied in \cite{Hui:2007cu,LoVerde:2007ke,Hui:2007tm,YOMI10}. In particular, the most recent work 
\cite{Vitto} demonstrated that the relativistic effects and wide-angle 
effects are of the same order and must be considered together. 
However, none of these previous studies are complete, when the relativistic
effects are concerned. In the presence of the gravitational potential 
contributions, the computation of the galaxy two-point correlation function
diverges in the infrared, a typical sign of theoretical deficiency, and 
as a consequence
one has to introduce an {\it arbitrary} cut-off scale~$k_{\rm IR}$ to the 
computation to keep the theoretical predictions under control.
A similar divergence in the infrared was observed in the variance of
the luminosity distance, and it was shown \cite{SG divergence,SG correlation}
that the such pathology appears due to the use of incorrect relativistic
descriptions.

Here we derive the two-point correlation function including all 
the relativistic effects in galaxy clustering.
The theoretical expression of  galaxy clustering used to compute the two-point correlation function must be gauge-invariant, as it represents a physical observable. By adopting a general metric representation with scalar and tensor perturbations we derive the relativistic expression of galaxy clustering (\cite{Yoo:2009,Yoo:2010,Bonvin:2011bg,Challinor:2011bk,Jeong}), showing its gauge invariance explicitly. In addition to the gauge invariance, the theoretical expression must be consistent with the equivalence principle of general relativity. Among other consequences, the latter implies that the physical observables are not affected by the (spatially) uniform gravity or ``long-mode'' perturbations.\footnote{Here we define the ``long-mode'' perturbations as those without any spatial variation over the scale of interest, so that their effect is to add uniform gravity (see sec.~\ref{EP} and in particular eqs.~\eqref{long-short} and \eqref{long-mode}).} It was shown in \cite{Jeong} that there is no such long-mode scalar contribution to galaxy clustering in the synchronous gauge, and this proof was generalized in \cite{cosmic rulers} for gravitational lensing. Drawing upon these studies, we demonstrate that our relativistic derivation of galaxy clustering is not affected by such long modes either from scalar or tensor perturbations. As described in \cite{SG divergence,YOGO15}, this also implies that our expression is devoid of infrared divergences. It is known that most relativistic expressions for galaxy clustering in literature have variances that diverge in the infrared \cite{Bertacca:2012tp}. This issue is usually ignored, and an arbitrary infrared cut-off is put in place to 
eliminate the divergences.
 Here we show how this issue can be naturally resolved, simply by using the correct theoretical expression. Furthermore, by showing that the gravitational potential contribution to the correlation function converges in the infrared, our study justifies an IR cut-off $(k_{\text{IR}} \leq H_0)$ in its computation. Indeed, the gravitational potential contribution is about 8 orders of magnitude smaller than the dominant one (the density fluctuation contribution) and cutting the integration at $k_{\text{IR}} \leq H_0$ does not bring any significant change in the total correlation function. On the other hand, we find that the contribution from the velocity at the observer is not relevant for the convergence of the correlation function in the infrared, but it cannot be neglected, as it is larger than the correlation of velocities at the sources.

With the correct theoretical prediction at hand, we numerically study the two-point correlation
function. 
Specifically, we derive the general analytic expressions for each relativistic effect in galaxy clustering: the density fluctuation, the redshift and the radial distortions, the gravitational lensing convergence and redshift-space distortions. This requires, in turn, to write down the correlation functions of the local potentials, the peculiar velocities, the integrated Sachs-Wolf effects and the Shapiro time-delay effects. 
Our study provides the amplitude of the correlation function for individual contributions, allowing to determine which effect dominates the total observed correlation in a given configuration of the galaxy pair. 
We perform the numerical investigation of the scalar perturbations in the conformal Newtonian gauge and the primordial gravitational-wave contributions. While the contribution to the observed angular galaxy clustering from gravitational waves has been studied already in \cite{DJFS LSS with tensors,DJFS LSS with tensors 2}, we generalize their results to the two-point galaxy correlation function.

The organization of the paper is as follows. 
First we study the galaxy number density theoretically in
sec.~\ref{sec Galaxy Clustering and Theoretical Investigations}, showing the gauge invariance in sec.~\ref{GI formalism} and the consistency with the equivalence principle in sec.~\ref{EP}. Then we study the two-point correlation function numerically in sec.~\ref{numerical}, where we first show that the correlation function does not exhibit infrared divergence. In sec.~\ref{scalar contribution} we analyze the correlation of individual relativistic effects, indicating the dominant contributions in different configurations. We perform the same analysis for the contribution of primordial gravitational waves in sec.~\ref{tensor contribution}.
We conclude with a summary and a discussion in sec.~\ref{discussion}.
In appendix \ref{app solutions} we provide the solution for the scalar perturbations needed for the numerical results of sec.~\ref{scalar contribution}.

\section{Galaxy Clustering and Theoretical Investigations}
\label{sec Galaxy Clustering and Theoretical Investigations}

In this section we derive the theoretical expression of the galaxy number density fluctuation to first order in perturbation theory. To prove the correctness of our expression we adopt a general metric representation and explicitly demonstrate the gauge-invariance of the theoretical expression. Then, in the conformal Newtonian gauge, we show that our expression is also consistent with the equivalence principle, further corroborating the sanity of our calculations.

\subsection{Metric convention and gauge transformations}
\label{metric}

Here we adopt a flat Friedmann-Robertson-Walker (FRW) metric with signature $(-,+,+,+)$ for our theoretical description of the background universe. 
In the presence of inhomogeneities, we parametrize the small perturbations to the background FRW metric by
\begin{equation}\label{metric p}
	\begin{split}
		 \delta g_{00} \equiv -2\, a^2 \alpha\,, \qquad \quad \delta g_{0i} \equiv -a^2 \beta_{,i}\,,
		\qquad \quad
		 \delta g_{ij} \equiv 2 \, a^2 \big[\varphi \, \bar g_{ij}+  \gamma_{,i|j} +C_{ij}\big]\,,
	\end{split}
\end{equation}
where $a$ is the scale factor, $\bar g_{ij}$ is the background 3-metric, commas represent the ordinary derivative while vertical bars represent the covariant derivative with $\bar g_{ij}$. The tensor perturbations $C_{ij}$ are constructed such that they are traceless $(C_{i}^{\,\,i}=0)$ and transverse $(C^{ij}_{\,\,\,\,\,\,|j}=0)$, with the longitudinal part being absorbed into the scalar perturbations.
The scalar ($\alpha, \beta, \varphi, \gamma$) and tensor ($C_{ij}$) perturbations are functions of a space-time point in a global coordinate $x^\mu=(\eta,\boldsymbol x)$, identified by a conformal time $\eta$ and spatial coordinates $x^i$, where Greek indices run over $0, 1, 2, 3$, while Latin indices over $1, 2, 3$. The metric representation in eq.~\eqref{metric p} is the most general accounting for scalar and tensor perturbations, and no gauge condition is imposed. In this paper we do not consider the vector perturbations, as they decay fast in time.
The observer motion is described by a time-like four-velocity $u^\mu \equiv a^{-1} \, (  1 - \alpha \,,\,\mathcal U^i\,)$, where the spatial component is further expressed in terms of a scalar perturbation $U$ as $\mathcal U^i \equiv - U^{,i}$. As we shall see in the next paragraph, it is convenient to define a scalar velocity $v \equiv U + \beta$, as it is independent of the spatial gauge transformation.

In order to obtain the gauge transformation properties of the metric perturbations introduced above we consider the coordinate transformation:
\begin{equation}\label{CT}
\tilde{x}^\mu=x^\mu+\xi^\mu \,, \qquad\qquad 	\xi^{\mu} \equiv (T,L^{,i}) \,,
\end{equation}
where the infinitesimal displacement field $\xi^\mu$ is decomposed in terms of two scalars $T$ and $L$.
The transformations of the metric perturbations are then given by \cite{Yoo, Bardeen,Dod}
\begin{equation}\label{gauge tr}
\begin{split}
&\tilde{\alpha} = \alpha - \frac{1}{a}(aT)' \,, \qquad\quad \tilde{\beta}=\beta-T+L'\,,  \qquad\quad \tilde{\gamma}=\gamma-L\,, 
\\
&\tilde{\varphi}=\varphi-\mathcal{H}T\,,
\qquad\qquad\,\, \tilde U = U - L'\,, \qquad\qquad\,\,\,\,\, \tilde v = v - T \,,
\end{split}
\end{equation}
where a prime indicates the derivative with respect to conformal time and $\mathcal H = a'/a = aH$ is the conformal Hubble parameter. Note that there is no gauge ambiguity for tensor perturbations at the linear order, $\tilde C_{ij}=C_{ij}$, as evident in eq.~\eqref{CT}. 
Based on the gauge transformation properties, we can define gauge-invariant quantities at linear level \cite{Yoo}:
\begin{equation}
	\alpha_\chi \equiv \alpha - \frac{1}{a}\chi'\,, \qquad \varphi_\chi \equiv \varphi-H\chi\,,  
	\qquad
	v_\chi \equiv v - \frac{1}{a}\chi\,, \qquad    \delta_v \equiv \delta + 3\,\mathcal H v\,,
\end{equation}
where $\chi \equiv a\,(\beta+\gamma')$ is the scalar shear of the normal observer, $n_\mu=-a(1+\alpha\,,\,0)$, transforming as $\tilde\chi=\chi-aT$, and $\delta$ is the matter density fluctuation, transforming as $\tilde \delta = \delta + 3\mathcal H T$. The notation for scalar gauge-invariant variables is set up such that $\alpha_\chi$, $\varphi_\chi$ and $v_\chi$ correspond respectively to the gravitational potentials and the velocity potential in the conformal Newtonian gauge ($\chi=0$), while $\delta_v$ is the matter density fluctuation in the comoving gauge ($v=0$) (see e.g. \cite{Ma,Dod}).
For convenience we define a gauge-invariant velocity $V^i \equiv - {v_\chi}^{,i}$ and a pure gauge term $\mathcal G^i \equiv \gamma^{,i}$ transforming as $\tilde{ \mathcal G}^i= \mathcal G^i -  L^{,i}$. As we re-arrange the perturbation variables in terms of gauge-invariant variables, we can easily isolate the gauge-dependent part that at linear order becomes $\mathcal G^i$.

\subsection{Gauge-invariant formalism of galaxy clustering}
\label{GI formalism}

In the past years, a number of groups have worked on the relativistic effects of galaxy clustering using the gauge-invariant formalism (\cite{Yoo:2009,Yoo:2010,Bonvin:2011bg,Challinor:2011bk,Jeong}, see also \cite{Yoo:2014sfa,Bertacca:2014dra,Bertacca:2014wga,DiDio:2015bua} for the second-order formalism).
The observed galaxy number density is obtained by counting the number of galaxies within the observed volume $dV_{\text{obs}}$ that appears to the observer as the volume within the observed redshift interval $dz$ and the observed solid angle $d\Omega$. In a homogeneous universe, the observed volume would be identical to the physical volume occupied by the observed galaxies.
However, in the presence of inhomogeneities in the universe, the observed volume $dV_{\text{obs}}$ does not correspond to the physical volume $dV \equiv dV_{\text{obs}}(1 + \delta V)$ and the difference is captured by the dimensionless fluctuation $\delta V$.
On the other hand, the number of observed galaxies $dN_g$ is unaffected by the inhomogeneities and can be expressed in terms of both the observed and the physical number densities, $n_g^{\text{obs}}$ and $n_g$, which are related by the volume fluctuation as
\begin{equation}
	dN_g \equiv n_g^{\text{obs}}\,dV_{\text{obs}} = n_g dV  \quad \longrightarrow \quad n_g^{\text{obs}} = n_g (1+\delta V)\,.
\end{equation}
In order to obtain the theoretical expression of the galaxy number density, we need to derive the fluctuation $\delta V$ in the volume occupied by the source galaxies. This requires the general relativistic relation of the observed redshift and angle to the physical volume along the past light-cone. 
Here we consider perturbations up to first order and we follow the gauge-invariant formalism developed in \cite{Yoo:2009,Yoo:2010,Yoo} to obtain the expressions of the volume fluctuation and then of the observed galaxy number density. Following \cite{Yoo}, we will first define the distortions in the position of source galaxies and subsequently use these to obtain the observables of our interest.

The position of a source galaxy is identified by the observed redshift $z$ and the observed angular position $\boldsymbol{\hat n}=(\sin\theta \cos\phi , \sin\theta \sin\phi , \cos\theta )$, measured in \textit{the observer rest frame}. Based on these quantities, the observer infers the source position $\bar x^\mu_s=(\bar\eta_z,\bar r_z \boldsymbol{\hat n})$ in a FRW coordinate by using the distance\,-\,redshift relation in a homogeneous universe,
\begin{equation}\label{d-z rel}
	\bar r_z = \bar \eta_o - \bar \eta_z = \int_0^z \frac{dz'}{H(z')}\,,
\end{equation}
where a bar denotes the coordinates at the observer ($o$) and the source (at redshift $z$) in the background.
The real position of the source is different from the inferred one, because the inhomogeneities affect the photon propagation. 
To account for the effect of the inhomogeneities on the real source position $x^\mu_s = (\eta_s,r_s,\theta_s,\phi_s)$ with respect to the inferred position $\bar x^\mu_s=(\bar\eta_z,\bar r_z , \theta, \phi)$ we define the time distortion $\Delta\eta \equiv \eta_s - \bar \eta_z$ (related to the distortion $\delta z$ in the observed redshift) and the geometric distortions of the spatial position $\delta r\equiv r_s - \bar r_z, \,\delta\theta \equiv \theta_s - \theta,\,\delta\phi \equiv \phi_s -\phi$.  

In this approach the redshift distortion and the time distortion are defined with respect to the observed redshift $1+z= 1/ a(\bar\eta_z)\equiv(1+\delta z)/a(\eta_s)$, which can be calculated as the ratio between the photon energy at the source and at the observer.\footnote{The photon energy is given by $E=-g_{\mu\nu}u^\mu k^\nu $, where $k^\mu$ is the photon wave-vector.} One obtains the following expression:
\begin{equation}
\delta z = \mathcal H \, \Delta\eta = -H\chi + ( \mathcal H\delta\eta + H\chi)_o + \big[V_i \hat n^i - \alpha_{\chi} \big]^z_o   
	  -  \int_0^{\bar r_z} d\bar r \, \big[\alpha_\chi - \varphi_\chi  - C_{ij}\hat n^i \hat n^j \big]' \,.
\end{equation}
The quantity $\delta\eta_o$ represents the observer time-lapse, describing the difference between the coordinate time at observation $\eta_o$ and the observer's proper time $\tau_o$. It is derived from the time component of the four-velocity $u^\mu = dx^\mu / d\tau$ as (see \cite{Yoo,Yoo:2016,Mine GLC,Yoo:2017svj} and in particular \cite{SG divergence} or \cite{Fanizza:2018qux} for a detailed derivation)
\begin{equation}
\delta\eta_o  = - \frac{1}{a_o} \int_0^{\bar\eta_o} d\bar\eta \, a \, \alpha \,,
\end{equation}
where $\bar \eta_o = \int_0^\infty \frac{dz}{H(z)}$ is uniquely determined and related to the observer proper time as $\tau_o=\int_0^{\bar \eta_o} d\eta \, a(\eta)$.
By making use of the gauge-invariant variables defined in sec.~\ref{metric}, the gauge-dependent term $-H\chi$ is isolated in the expression of $\delta z$, which transforms in fact as $\widetilde{\delta z} = \delta z + \mathcal H T$. We can therefore define a new gauge-invariant variable $\delta z_\chi = \delta z + H \chi$.

The geometric distortions of the source spatial position $\delta x^i_s \equiv x^i_s - \bar x^i_s$ can be computed by integrating the photon geodesic equation from the observer position to the source position, as described in \cite{Yoo:2009,Yoo:2010,Yoo}. By following that approach we obtain
\begin{equation}
	\delta r = \hat n_i\, x^i_s - \bar r_z = - \hat n_i \mathcal G^i + \hat n_i (\delta x^i + \mathcal G^i )_o  +  (\delta \eta + \chi)_o - \frac{\delta z_\chi}{\mathcal{H}_z}
 + \int_0^{\bar r_z} d\bar r\,[\alpha_\chi-\varphi_\chi  - C_{ij}\hat n^i \hat n^j]   \,,
\end{equation}
\begin{equation}
	\begin{split}
		\bar r_z \delta\theta = \hat\theta_i\, x^i_s &= -\hat\theta_i \mathcal G^i + \hat\theta_i (\delta x^i + \mathcal G^i)_o + \bar r_z \hat\theta_i \big( -  V^i  + C^i_j \hat n^j \big)_o  
	\\
		& \quad  - 2 \int_0^{\bar r_z}d\bar r\, \hat\theta_i  C^i_j  \hat n^j 
		  -\int_0^{\bar r_z}d\bar r\,  (\bar r_z - \bar r) \hat\theta_i [(\alpha_\chi -\varphi_\chi)^{,i} - {C_{jk}}^{,i}\hat n^j \hat n^k ]		\,, \qquad\qquad
	\end{split}
\end{equation}	
\begin{equation}
	\begin{split}
		\bar r_z\sin\theta\, \delta\phi = \hat\phi_i\, x^i_s &= -\hat\phi_i \mathcal G^i + \hat\phi_i (\delta x^i + \mathcal G^i)_o + \bar r_z \sin\theta\,  \hat\phi_i \big( -  V^i  + C^i_j \hat n^j \big)_o  
	\\
	  &\quad -2 \int_0^{\bar r_z}d\bar r\, \hat\phi_i C^i_j  \hat n^j 
		  -\int_0^{\bar r_z}d\bar r\,  (\bar r_z - \bar r) \hat\phi_i [(\alpha_\chi -\varphi_\chi)^{,i} - {C_{jk}}^{,i}\hat n^j \hat n^k ]		\,, \quad
	\end{split}
\end{equation}
where the unit vectors $\hat\theta^i = \partial_\theta \hat n^i$ and $\hat\phi^i = (1/\sin\theta) \, \partial_\phi \hat n^i$ are projectors on the sphere.
The quantity $\delta x^i_o$ represents the spatial shift at the observer position, describing the change caused by the velocity field generated by the inhomogeneities. Exactly in the same way as the observer coordinate lapse $\delta\eta_o$, it is derived by integrating the spatial part of the four-velocity  as (see \cite{Fanizza:2018qux,Yoo:2017svj} for a detailed derivation)
\begin{equation}
\delta x^i_o =  - \int_0^{\bar\eta_o} d\bar\eta \,\,  U^{,i} \,.
\end{equation}
This effect has been often neglected in the literature, but, as we shall see, it cancels out in any linear-order expression of the observables. 

Given the angular distortions above, one can compute also the change in the solid angle subtended by the source. 
This effect is known as the gravitational lensing convergence and is given by the ratio between the observed solid angle and the solid angle at the source as
\begin{equation}
	\begin{split}
		\kappa &\equiv -\frac{1}{2} \bigg[\bigg( \cot\theta +\frac{\partial}{\partial\theta}\bigg)\delta\theta + \frac{\partial}{\partial\phi}\delta\phi \bigg]
		\\
		&= - \frac{\hat n_i \mathcal G^i}{\bar r_z} + \frac{1}{2\bar r_z} \hat\nabla_i  \mathcal G^i  +    \frac{\hat n_i}{\bar r_z} (\mathcal G^i + \delta x^i)_o + \hat n_i \bigg( - V^i + \frac{3}{2} C^i_j \hat n^j  \bigg)_o 
		\\
		 & \quad  
		 - 2 \int_0^{\bar r_z}d\bar r\, \frac{  C_{ij}\hat n^i \hat n^j }{\bar r_z}
		 +  \int_0^{\bar r_z}d\bar r\,  \frac{\hat\nabla_i  ( C^i_j \, \hat n^j )}{\bar r_z} 
		 \\
		 & \quad
		 +\int_0^{\bar r_z}d\bar r\, \bigg(\frac{\bar r_z - \bar r}{2\,\bar r_z \bar r}\bigg)\big[ \hat\nabla^2 (\alpha_\chi -\varphi_\chi) - \big( \hat n^i \hat n^j \hat\nabla^2 C_{ij} + 2\,\hat n^i \hat\nabla_j C^j_i \big)\big] \,,
	\end{split}
\end{equation}
where $\hat \nabla_i$ is the angular gradient operator and $\hat\nabla^2$ is the angular Laplacian. 
The gauge transformation properties are transparent: 
\begin{equation}
\begin{split}
	&\widetilde{\delta r} = \delta r + \hat n_i L^{,i}, \qquad \bar r_z \widetilde{\delta\theta} = \bar r_z \delta\theta + \hat\theta_i L^{,i} \,, \qquad \bar r_z\sin\theta\, \widetilde{\delta\phi} = \bar r_z\sin\theta\, \delta\phi + \hat\phi_i L^{,i} \,,
		\\
	&	\tilde\kappa = \kappa + \frac{\hat n_i  L^{,i}}{\bar r_z} - \frac{1}{2\,\bar r_z}\hat \nabla_i  L^{,i} \,,
\end{split}
\end{equation}
and this shows that the real position $x^\mu_s$ of the source is a coordinate-dependent quantity.
As for the redshift distortion $\delta z_\chi$, the expressions of $\delta r$, $\delta\theta$, $\delta\phi$ and $\kappa$ can be arranged in terms of gauge-invariant variables, isolating the gauge-dependent terms (involving $\mathcal G^i$), as 
\begin{equation}
	\delta r_\chi = \delta r + \hat n_i \mathcal G^i, \qquad\qquad \mathcal K = \kappa + \frac{\hat n_i \mathcal G^i}{\bar r_z} - \frac{1}{2\,\bar r_z}\hat \nabla_i \mathcal G^i \,.
\end{equation}
Since the effects of the inhomogeneities are conveniently expressed in terms of the geometric distortions that we have introduced, we can write explicitly gauge-invariant expressions of the cosmological observables. 

Now we use the gauge-invariant formalism summarized above (see \cite{Yoo} for the extensive description) to derive first the fluctuation in the luminosity distance and then that in the galaxy number density.
The fluctuation $\delta \mathcal D_L$ in the luminosity distance is defined through $\mathcal D_L \equiv \bar{\mathcal D}_L (1 + \delta \mathcal D_L)$, where $\bar{\mathcal D}_L = (1+z)\bar r_z$. From its exact relation with the angular diameter distance $\mathcal D_A = (1+z)^{-2} \mathcal D_L$, we can compute $\delta \mathcal D_A$ with ease, by using the geometric distortions for a unit area. The angular diameter distance is the distance at which a solid angle $d\Omega=\sin\theta d\theta d\phi$ subtends a physical area $dA$ perpendicular to the photon propagation in the source rest frame,
\begin{equation}
	dA \equiv \mathcal D_A^2 d\Omega = \sqrt{-g} \, \epsilon_{\mu\nu\rho\sigma}u_s^\mu n^\mu_s \frac{\partial x^\rho_s}{\partial\theta}\frac{\partial x^\sigma_s}{\partial\phi}d\theta d\phi \,,
\end{equation}
where $n^\mu = k^\mu/(k^\nu u_\nu)+u^\mu$ is the observed photon direction for the observer with four-velocity $u^\mu$.
From this equation we obtain the fluctuation in the distance as a function of the observed redshift and angles
\begin{equation}
	\delta \mathcal D_L (z,\boldsymbol{\hat n}) = \delta \mathcal D_A =  \delta z_\chi + \frac{\delta r_\chi}{\bar r_z} - \mathcal K + \varphi_\chi - \frac{1}{2} C_{ij}  \hat n^i \hat n^j  .
\end{equation} 
Written in terms of gauge-invariant variables, the gauge-invariance of the luminosity distance fluctuation is manifest (see \cite{Yoo,SG divergence,Mine GLC}). 
Indeed, the luminosity distance is an observable, here expressed in terms of the other observables (redshift and angles), and therefore must be independent from the gauge conditions chosen \cite{Bardeen}. Note the cancellation of the observer spatial shift $\delta x^i_o$ among the radial distortion and the lensing convergence. As anticipated, this occurs for the expression of any observable at linear level.

By extending the previous expression of the infinitesimal area in the source rest frame, the infinitesimal volume occupied by the source galaxies is given by
\begin{equation}
	dV = \sqrt{-g} \, \epsilon_{\mu\nu\rho\sigma}u_s^\mu \frac{\partial x^\nu_s}{\partial z}  \frac{\partial x^\rho_s}{\partial\theta}\frac{\partial x^\sigma_s}{\partial\phi}dz d\theta d\phi \equiv dV_{\text{obs}} (1+\delta V) \,, \qquad dV_{\text{obs}} =  \frac{\bar r_z^2 dz d\Omega}{H(1+z)^{3}} \,.
\end{equation}
Thus, one obtains the linear-order relativistic correction to the physical volume
\begin{equation}\label{delta V}
	\delta V = 3\,\delta z_\chi + 3\,\varphi_\chi + 2\, \frac{\delta r_\chi}{\bar r_z} - 2\,\mathcal K + H \frac{\partial}{\partial z} \delta r_\chi + V_i \hat n^i \,,
\end{equation}
which is manifestly gauge-invariant, as required by the fact that the volume itself is an observable.  
Finally, we have all ingredients to get the galaxy number density and its fluctuation.
We can write the observed and physical galaxy number densities respectively as
\begin{equation}\label{ud}
	n_g^{\text{obs}} \equiv \bar n_g (\bar \eta_z) (1+\delta_g^\text{obs}) \,, \qquad\qquad n_g \equiv \bar n_g ( \tau_s) (1+\delta_g^{\text{int}}) \,,
\end{equation}
where we have defined the fluctuations $\delta_g^\text{obs}$ and $\delta_g^{\text{int}}$.
Note that the mean density $\bar n_g$ and the intrinsic fluctuation $\delta_g^\text{int}$ in the physical density are defined over the proper-time hypersurface of the source described by the comoving-synchronous gauge.  By denoting the proper-time hypersurface with $\tau_s$, the intrinsic fluctuation can be written as $\delta_g^{\text{int}} \equiv b \, \delta^{\tau_s}_m \equiv b \, \delta_v $, where $b$ is the galaxy bias and $\delta^{\tau_s}_m \equiv  \delta_v$ is the matter density fluctuation in the comoving-synchronous gauge \textcolor{red}{\cite{Jeong,Yoo}}.
Thus, the observed galaxy number density fluctuation is given by
\begin{equation}\label{d g}
	\delta_g^\text{obs}(z,\boldsymbol{\hat n}) =  b \, \delta_v -e_z\, \delta z_{v} +  \delta V \,, \qquad\qquad e_z \equiv - \frac{1}{\mathcal H_z}\frac{\bar n_g' ( \bar\eta_z)}{\bar n_g (\bar \eta_z)}=\frac{d\ln \bar n_g}{d\ln (1+z)} \,.
\end{equation}
Any quantity in the above expression is \textit{gauge-invariant}, indeed $\delta_v$ and $\delta z_v$ are those in the comoving-synchronous gauge and the gauge-invariance of the volume distortion is explicitly verified by expressing it in terms of gauge-invariant variables as in eq.~\eqref{delta V}.

\subsection{Compatibility check with the equivalence principle}
\label{EP}

Following the lead by \cite{Jeong,cosmic rulers,SG divergence}, we perform the compatibility check of our theoretical expression with the equivalence principle.
The gauge invariance and the equivalence principle of general relativity offer a powerful way to test the validity of our theoretical predictions in sec.~\ref{GI formalism}. The gauge-invariance reflects the fact that the physics is independent of the way the perturbations are defined with respect to the fictitious background. 
The equivalence principle asserts the physical equivalence of a gravitational field and its corresponding acceleration of the reference system.
It implies that the laws of physics in a reference frame that is in free fall are the same as in the complete absence of gravity, i.e. the laws of physics are those of special relativity.
Strictly speaking, however, the equivalence principle is applicable to the limit in which the differential gravity, or the tidal force can be neglected.
The tidal effects are, indeed, the leading physical effect of gravity.
Applying the equivalence principle to the case of our interest, where the source and the observer are on the past light-cone with the unique scale set by the (comoving) distance $\bar r_z$, we will consider only the long-mode perturbations that are spatially uniform over the scale $\bar r_z$ and show that galaxy clustering is independent of such long-mode perturbations.

In the previous subsection we showed that our expressions of the luminosity distance and galaxy number density are gauge-invariant. In this subsection we further check the compatibility of these expressions with the equivalence principle. 
According to the latter, as discussed above, the uniform gravity generated by long-mode perturbations should have no consequence on the physical observables. We will isolate in the perturbations the contributions to a (spatially) uniform gravitational field and show that our expressions are devoid of these terms.
Besides confirming our derivations, we show that our expressions do not exhibit any infrared divergence on super horizon-scales, as demonstrated in \cite{SG divergence,YOGO15}. 

To focus on the effects of such long-mode perturbations we take the Fourier transformation of the perturbation variables and introduce a cut-off scale $k_{\text{IR}}$ set by $k_{\text{IR}} \bar r_z  \ll 1$. 
To elaborate on this, let us consider a gravitational potential $\Psi(\eta,\boldsymbol{x})$ and its Fourier mode $\Psi(\eta,\boldsymbol{k})$, where at the source $|\boldsymbol x | = \bar r_z$ and $\boldsymbol x / \bar r_z = \boldsymbol {\hat n}$. The gravitational potential can be split into the long-mode and short-mode contributions as
\begin{equation}\label{long-short}
	\Psi(\eta,\boldsymbol{x})= \bigg( \int_0^{k_{\text{IR}}} + \int_{k_{\text{IR}}}^{\infty} \bigg) \frac{d^3 k}{(2\pi)^3}\, e^{i\, \boldsymbol{k}\cdot\boldsymbol{x}}\,  \Psi(\eta,\boldsymbol{k}) \equiv \Psi^\ell(\eta,\boldsymbol{x}) + \Psi^s(\eta,\boldsymbol{x})  \,.
\end{equation}
By expanding in terms of $k \,  x \, ( \leq k_{\text{IR}} \bar r_z \ll 1 )$, the long-mode potential can be written as 
\begin{equation}\label{long-mode}
\begin{split}
	\Psi^\ell(\eta,\boldsymbol{x}) &=  \int_0^{k_{\text{IR}}} \frac{d^3 k}{(2\pi)^3}\, \bigg( 1 + i \boldsymbol{k}\cdot\boldsymbol{x} - \frac{1}{2}(\boldsymbol{k}\cdot\boldsymbol{x})^2 + \ldots \bigg)\,  \Psi(\eta,\boldsymbol{k})
	\\
	&= \Psi^\ell_o(\eta) + x^i \big[\partial_i \Psi^\ell \big]_o(\eta) + \frac{1}{2} x^i x^j \big[\partial_i \partial_j \Psi^\ell\big]_o(\eta) + \ldots \,,
\end{split}
\end{equation}
where we defined several functions
\begin{equation}
	\Psi_o^\ell(\eta)  \equiv \int_0^{k_{\text{IR}}}  \frac{d^3 k}{(2\pi)^3} \Psi(\eta,\boldsymbol{k}) \,, \qquad  [\partial_i \cdots \partial_j \Psi^\ell]_o (\eta) \equiv \int_0^{k_{\text{IR}}}  \frac{d^3 k}{(2\pi)^3}\, (i k_i)\cdots (i k_j)   \Psi(\eta,
\boldsymbol{k})\,, 
\end{equation}
evaluated spatially at the origin $\boldsymbol{x}=0$.
With these definitions, the first term $\Psi_o^\ell$ represents the contribution of the \textit{uniform gravitational potential} to $\Psi(\eta,\boldsymbol{x})$, while the second term $x^i [\partial_i \Psi^\ell]_o$ represents the contribution of the \textit{uniform gravitational force}.  
According to the equivalence principle, both $\Psi^\ell_o$ and $x^i [\partial_i \Psi^\ell]_o$ should have no effect on physical observables, as their contributions are indistinguishable from the free-fall.  On the other hand, the third term in eq.~\eqref{long-mode} is relevant, as it is responsible for tidal effects. This concept will be generalized to tensor perturbations.

We are now going to show that our theoretical expressions for the luminosity distance and the galaxy number density do not contain the terms discussed above.
Since the full expressions in sec.~\ref{GI formalism} are gauge-invariant, we choose the conformal Newtonian gauge for simplicity to demonstrate the compatibility with the equivalence principle.
As we assume no anisotropic stress and no vector perturbations in the universe, our metric is given by
\begin{equation}
\label{metric NG}
	ds^2 = -a^2 (1+2\,\Psi)d\eta^2 + a^2 \big[(1 - 2\,\Psi)\bar g_{ij} + 2\,C_{ij}\big]dx^i dx^j \,, \qquad \mathcal G^i = 0 \,,
\end{equation}
where we have denoted the gravitational potential as $\alpha_\chi = - \varphi_\chi \equiv \Psi$.
Having removed any gauge ambiguity, we will simply drop the subscript $\chi$ in the other variables defined in secs.~\ref{metric p} and \ref{GI formalism}.

\subsubsection{Scalar perturbations}
\label{sec: scalar}

We first consider only the scalar perturbations.
In the conformal Newtonian gauge with only scalar perturbations the expressions of the luminosity distance, the volume and the galaxy number density fluctuations are 
\begin{equation}\label{DLS-dgNG-dV}
\begin{split}
	&\delta \mathcal D_L  =  \delta z + \frac{\delta r}{\bar r_z} - \mathcal K - \Psi \,, \qquad \delta V =  3\, \delta z  - 3\,\Psi + 2\, \frac{\delta r}{\bar r_z} - 2\,\mathcal K + H_z \frac{\partial}{\partial z} \delta r - \hat n^i {v}_{,i}\,,
\\
	& \delta_g = (b \, \delta_v -e_z \, \delta z_v)   + \delta V .
\end{split}
\end{equation}
The geometric distortions are given in terms of the scalar potentials for gravity $\Psi$ and velocity $v$ by
\begin{equation}
\begin{split}
	&\delta z =  \mathcal H_o {\delta\eta}_o - \big[ \hat n^i {v}_{,i} + \Psi \big]^z_o   
	  -  2 \int_0^{\bar r_z} d\bar r \, {\Psi}' \,, 
\\
	&\delta r =  \hat n_i \, \delta x^{i}_o  +  {\delta\eta}_o - \frac{\delta z}{\mathcal{H}_z}
	 + 2 \int_0^{\bar r_z} d\bar r\,\Psi \,,
\\
	& \mathcal K =  \frac{\hat n_i \, \delta x^{i}_o}{\bar r_z}  + \big( \hat n^i {v}_{,i}\big)_o 
		 +\int_0^{\bar r_z}d\bar r\, \bigg(\frac{\bar r_z - \bar r}{\bar r_z \bar r}\bigg)\hat\nabla^2 \Psi \,,
		 \end{split}
\end{equation}
where the coordinate lapses at the observer are related to the velocity potential $v$ as 
\begin{equation}
	\delta \eta_o = - v_o \,, \qquad\qquad  \delta x^{i}_{o}=-\int_0^{\bar \eta_o}d\bar \eta\, {v}^{\,,i}(\bar\eta, \boldsymbol{x}_o)\,.
\end{equation}
As described in appendix \ref{app solutions}, at linear order we can separate the gravitational potential $\Psi(\eta,\boldsymbol{x})$ in terms of the growth function $D_\Psi(\eta)$ and the curvature perturbation $\zeta(\boldsymbol{x})$ in the comoving gauge: $\Psi(\eta,\boldsymbol{x})=D_\Psi(\eta)\zeta(\boldsymbol{x})$. The curvature perturbation $\zeta(\boldsymbol{x})$ is constant in time and related to the growing mode $\delta_+(\boldsymbol{x})$ of the density contrast $\delta(\eta,\boldsymbol{x})\equiv D(\eta)\delta_+(\boldsymbol{x})$. Accordingly, the gravitational potential growth function $D_\Psi(\eta)$ is related to the matter growth function $D(\eta)$, whose solution is given in eq.~\eqref{D1}.
The long-mode gravitational potential is then proportional to the long-mode curvature perturbation and can be expanded as in eq.~\eqref{long-mode},
\begin{equation}\label{zeta}
\Psi^{\ell}(\eta,\bar r\,\boldsymbol{\hat n})=D_\Psi(\eta)\zeta^\ell(\bar r\,\boldsymbol{\hat n}) = 	D_\Psi(\eta) \big[ \zeta_o +  \bar r\,\zeta_1 (\boldsymbol{\hat n})  + \ldots \big],
\end{equation}
where we have defined 
\begin{equation}
	\zeta_o \equiv \zeta^{\ell} \big\vert_o = \int_0^{k_{\text{IR}}}  \frac{d^3 k}{(2\pi)^3} \zeta(\boldsymbol{k}) \,, \qquad\qquad 
	\zeta_1 (\boldsymbol{\hat n}) \equiv  \hat n^i \big[ \partial_i \zeta^{\ell} \big]_o = \hat n^i \int_0^{k_{\text{IR}}}  \frac{d^3 k}{(2\pi)^3}\, i k_i \, \zeta(\boldsymbol{k}) \,.
\end{equation}
Analogously, the long-mode velocity potential is given by
\begin{equation}\label{v pot}
	v^\ell(\eta,\bar r\,\boldsymbol{\hat n}) = - D_V(\eta)\zeta^\ell (\bar r\,\boldsymbol{\hat n}) = - D_V(\eta)\big[ \zeta_o +  \bar r\,\zeta_1(\boldsymbol{\hat n}) + \ldots \big] \,,
\end{equation}
where the dimension of $v$ and $D_V$ is $[v]=[D_V]=L$ and $D_V$ is related to $D_\Psi$ through the Einstein equations, as derived in appendix \ref{app solutions}. In particular, the following relations are essential for our purpose: 
\begin{equation}\label{rel}
	 D_\Psi=\mathcal H D_V -1  = - \frac{1}{2}(D_V'+1) \,,  \qquad  \int_0^{\bar r_z}d\bar r\, D_\Psi = \frac{1}{2}(D_V - D_{Vo} - \bar r_z) \,.
\end{equation}
Now we demonstrate that our theoretical expressions for the luminosity distance, the volume and the galaxy number density fluctuations are independent of the uniform gravitational field generated by $\zeta_o$ and the uniform acceleration field generated by $\zeta_1$.

In the long-mode limit, where the wavelength of perturbations is much larger than the distance between the observer and the source $(k_{\text{IR}} \bar r_z  \ll 1)$, we take the potentials as $\Psi \equiv \Psi^\ell \equiv D_\Psi ( \zeta_o +  \bar r_z\, \zeta_1 )$ and $v \equiv v^\ell  \equiv - D_V ( \zeta_o + \bar r_z\, \zeta_1 )$.
The geometric distortions in terms of $\zeta_o$ and $\zeta_1$ are then 
\begin{equation}\label{scalar mon dip}
\begin{split}
	& \delta z(\zeta_o,\zeta_1 ) =  [ D_\Psi +1 ] ( \zeta_o + \bar r_z \,  \zeta_1  ) \,,
	\\ 
	& \delta r(\zeta_o,\zeta_1 ) =  \hat n_i\, \delta x^{i}_o (\zeta_1 ) -\bar r_z \, \zeta_o - \frac{1}{\mathcal{H}_z}[ D_\Psi + 1 ] \bar r_z \, \zeta_1 + 2 \, \zeta_1  \int_0^{\bar r_z} d\bar r\, \bar r \,D_\Psi \,, 
	\\
	& \mathcal K (\zeta_o,\zeta_1 ) =  \frac{\hat n_i\, \delta x^{i}_o(\zeta_1 )}{\bar r_z}  -   \frac{1}{\mathcal H_z}[D_\Psi + 1] \zeta_1  + \bar r_z \, \zeta_1 + 2\, \zeta_1 \int_0^{\bar r_z}d\bar r \, \frac{\bar r}{\bar r_z}D_\Psi\,,
\end{split}
\end{equation}
where we have used eq.~\eqref{rel} to express the time dependence only through $D_\Psi$ (and not $D_V$). Note that the lensing convergence is only affected by $\zeta_1(\boldsymbol{\hat n})$ but not $\zeta_o$, while the redshift and the radial distortions contain both terms. This is explained by the fact that $\mathcal K$ describes only transverse effects with respect to the line of sight $\boldsymbol{\hat n}$ and a constant scalar like $\zeta_o$ has no transverse components. On the other hand, the uniform acceleration associated with $\zeta_1(\boldsymbol{\hat n})$ generates a velocity that inevitably affects the convergence~$\mathcal K$, as the observed solid angle changes.
By substituting the above contributions into $\delta \mathcal D_L$ and $\delta V$ as in eq.~\eqref{DLS-dgNG-dV} we easily verify that the scalar expression of the luminosity distance and the volume are not affected by the uniform gravity generated by long-mode scalar perturbations,
\begin{equation}
	\delta \mathcal D_L(\zeta_o,\zeta_1 ) = 0 \,, \qquad\qquad
	\delta V (\zeta_o,\zeta_1 ) = 0 \,,
\end{equation} 
in agreement with the equivalence principle. 

Now, to show that $\delta_g$ is likewise not affected by the uniform gravity we only need to prove that $\delta_v(\zeta_o,\zeta_1 )=\delta z_v(\zeta_o,\zeta_1 )=0$, as $b \neq e_z$ in general. 
First of all, the matter density fluctuation $\delta_v$ in the comoving gauge is not affected by uniform gravity because the Einstein equation dictates $\delta_v \propto \Delta \Psi$. 
To prove that also $\delta z_v(\zeta_o,\zeta_1 )=0$ we first need to transform the redshift distortion from the comoving gauge to the conformal Newtonian gauge. By considering the gauge transformations of $\beta$ and $\gamma$ in eq.~\eqref{gauge tr} we obtain that the displacement field $\xi^\mu$ in eq.~\eqref{CT}, which generates the transformation from the comoving gauge $(\gamma=v=0)$ to the conformal Newtonian gauge $(\beta=\gamma=0)$, is given by $T=\beta$ and $L=0$. Then, from the gauge transformations of $v$ and $\delta z$ we have that $\beta = - v$ and 
$\delta z_v =\delta z + \mathcal H \, v $. At this point it is straightforward to verify that  $\delta z_v(\zeta_o,\zeta_1 ) = [ D_\Psi +1 - \mathcal H D_V ] ( \zeta_o + \bar r_z \,  \zeta_1  )=0$, because from eq.~\eqref{rel} we have that $\mathcal H D_V = D_\Psi + 1$. 
The fact that the redshift distortion in the comoving-synchronous gauge is devoid of the long-mode contributions can also be readily understood as follows. The redshift $z$ is a gauge-invariant physical observable but the redshift distortion $\delta z$ is not, as it compensates the difference between the  time of photon emission in a homogeneous universe $\bar\eta_z$ and the true coordinate time at the source $\eta_s$, which changes from one gauge to another.
However, in the comoving-synchronous gauge the degrees of freedom in the perturbations are fixed such that at the observer the physical space-time corresponds to the background, i.e. the lapse functions are vanishing. Consequently, a redshift measurement would provide unambiguous information (independent from the potentials at $o$) about the emission time of the photons. This time measurement cannot be influenced by uniform gravity. In turn, the redshift distortion in the comoving gauge has to be unaffected by uniform gravity, as there is no mode to be compensated.
We conclude that the expression of the galaxy number density fluctuation is free from the uniform gravitational potential and accelaration contributions
\begin{equation}
	\delta_g(\zeta_o,\zeta_1 )=0 \,.
\end{equation} 
Being independent from the presence of a uniform gravitational field, our expression is compatible with the equivalence principle.

\subsubsection{Tensor perturbations}
\label{sec: tensor}

We now demonstrate that the luminosity distance and the galaxy number density are not affected by the uniform gravity generated by long-mode tensor perturbations from inflation.
The expressions of these observables when only tensor perturbations are taken into account are 
\begin{equation}\label{DLT}
	\delta \mathcal D_L  = \delta z + \frac{\delta r}{\bar r_z} - \kappa  - \frac{1}{2} C_{ij}  \hat n^i \hat n^j  \,, \qquad \delta_g = (3-e_z)\delta z  + 2\, \frac{\delta r}{\bar r_z} - 2\,\kappa + H \frac{\partial}{\partial z} \delta r \,,
\end{equation}
where the geometric distortions are given in terms of the tensor perturbations $C_{ij}$ by
\begin{equation}\label{kappa tens}
\begin{split}
\delta z &=   \int_0^{\bar r_z} d\bar r \,  {C_{ij}}'\hat n^i \hat n^j  \,,
\qquad\qquad
	\delta r = - \frac{\delta z}{\mathcal H_z} -  \int_0^{\bar r_z} d\bar r\, C_{ij}\hat n^i \hat n^j  \,, \qquad\qquad \delta\eta_o = \delta x^i_o = 0 \,,
\\
		\kappa &= \frac{3}{2} \big( C_{ij} \hat n^i \hat n^j  \big)_o   
		 - \int_0^{\bar r_z}d\bar r\, \frac{   2C_{ij} \hat n^i \hat n^j - \hat\nabla_i  (C^i_j  \hat n^j )}{\bar r_z} 
		  - \int_0^{\bar r_z}d\bar r\, \bigg(\frac{\bar r_z - \bar r}{2\,\bar r_z \bar r}\bigg)\big[  \hat n^i \hat n^j \hat\nabla^2 C_{ij} + 2\hat n^i \hat\nabla_j C^j_i \big]  \,.
	\end{split}
\end{equation}
Tensor perturbations can be decomposed into Fourier modes of two independent polarization states labeled as $s=+,\times$,
\begin{equation}\label{tensor dec}
	C_{ij}(\eta,\boldsymbol{k})= e^+_{ij}(\boldsymbol{\hat k})\, C_+(\eta,\boldsymbol{k})+e^\times_{ij}(\boldsymbol{\hat k})\,C_\times(\eta,\boldsymbol{k}) \,,
\end{equation}
where the basis tensors $e^{s}_{ij}(\boldsymbol{\hat k})$ are transverse, traceless and normalized through $e^s_{ij}e^{s'ij} \equiv 2\,\delta^{ss'}$. 
Using the Einstein equation in Fourier space in the absence of anisotropic pressure,
\begin{equation}
	C_s''(\eta,\boldsymbol{k}) + 2\, \mathcal H \, C_s'(\eta,\boldsymbol{k}) + k^2 C_s(\eta,\boldsymbol{k}) = 0\,,
\end{equation}
we find that, considering long-mode perturbations $(\text{for } k^2 \approx 0)$ and neglecting decaying modes in the solution, each polarization $C_s$ of the tensor perturbations is constant in time, i.e. ${C_{s}^{\ell}}' = 0$.
In real space the long-mode primordial gravitational waves can then be written as
\begin{equation}
\begin{split}
C^\ell_{ij}(\bar r \, \boldsymbol{\hat n}) &= \int_0^{k_{\text{IR}}} \frac{d^3k}{(2\pi)^3}e^{i \bar r\, \boldsymbol{\hat n}\cdot \boldsymbol{k} }\,e^s_{ij}(\boldsymbol{\hat k})\, C_{s}(\boldsymbol{k})
		\\
		&= \int_0^{k_{\text{IR}}} \frac{d^3k}{(2\pi)^3} \big[ 1 + i \bar r\, \boldsymbol{\hat n}\cdot \boldsymbol{k}  + \ldots \big]\,e^s_{ij}(\boldsymbol{\hat k})\,C_{s}(\boldsymbol{k})
		\\
	&=  C_{ijo} + \bar r\,C_{ij1}(\boldsymbol{\hat n}) + \ldots \,,
\end{split}
\end{equation}
where we have defined 
\begin{equation}
	C_{ijo} \equiv C^\ell_{ij} \big\vert_o = \int_0^{k_{\text{IR}}} \frac{d^3k}{(2\pi)^3} e^s_{ij}(\boldsymbol{\hat k})C_{s}(\boldsymbol{k}) \,, \quad  C_{ij1}(\boldsymbol{\hat n}) \equiv  \hat n^k\big[ \partial_k C^\ell_{ij} \big]_o = \hat n^k \int_0^{k_{\text{IR}}} \frac{d^3k}{(2\pi)^3} i k_k \, e^s_{ij}(\boldsymbol{\hat k}) C_{s}(\boldsymbol{k}).
\end{equation}

We start again by studying the contributions of the long-modes $C_{ijo}$ and $C_{ij1}$ to the individual components in the luminosity distance and the galaxy number density. In the long-mode limit, where the perturbations wavelength is much larger than the scale of our system $(k_{\text{IR}} \bar r_z  \ll 1)$, we take the gravitational waves as $C_{ij} \equiv C_{ij}^\ell \equiv  C_{ijo} + \bar r_z\,C_{ij1} $.
The geometric distortions in terms of $C_{ijo}$ and $C_{ij1}$ are then 
\begin{equation}
\label{tensor first}
\begin{split}
& \delta z(C_{ijo},C_{ij1}) =   0  \,,
\qquad\qquad
\delta r(C_{ijo},C_{ij1}) =  -  \bar r_z\, C_{ijo}\hat n^i \hat n^j -  \frac{1}{2}\bar r_z^2\, C_{ij1} \hat n^i \hat n^j  \,,
\\
& \kappa(C_{ijo},C_{ij1}) = -\frac{3}{2}  C_{ijo} \hat n^i \hat n^j -  \bar r_z\,C_{ij1} \hat n^i \hat n^j   \,.
\end{split}
\end{equation}
By substituting these expressions into eq.~\eqref{DLT} we verify straightforwardly that the luminosity distance, the volume distortion, and the galaxy number density are not affected by the long-mode primordial gravitational waves,
\begin{equation}
\delta \mathcal D_L(C_{ijo},C_{ij1})=\delta V(C_{ijo},C_{ij1})=
 \delta_g(C_{ijo},C_{ij1}) = 0 \,.
\end{equation}
As a conclusion, our theoretical expressions for the luminosity distance and the galaxy number density are independent from the presence of a uniform gravitational field and, therefore, consistent with the equivalence principle.

\section{Numerical Investigation of the Galaxy Two-Point Correlation Function}
\label{numerical}

Galaxy clustering is a key observable in cosmology and constitutes the main subject of our study. In particular, the two-point correlation function $\langle \delta_g(\boldsymbol{x})\delta_g(\boldsymbol{x}+\boldsymbol{r})\rangle$ measures the excess of probability of finding a pair of galaxies separated by a vector $\boldsymbol{r}$ relative to the uniform distribution $\bar n_g(z)$ in eq.~\eqref{ud}. Of course, the two-point statistics is affected by the same relativistic effects altering the observed galaxy number density. In this section, we compute numerically the two-point correlation functions of the various contributions to the linear-order fluctuation $\delta_g$. These contributions are the matter density contrast $\delta_v$, the redshift and radial distortions $\delta z$ and $\delta r$, the gravitational lensing convergence $\mathcal K$ and the term $H_z \frac{\partial}{\partial z} \delta r$, which includes the so-called Kaiser effect (or redshift space distortion), as we shall see.

As in the previous sections, we neglect the vector perturbations and we consider scalar and tensor perturbations separately. Again, we consider the conformal Newtonian gauge with metric given in eq.~\eqref{metric NG}. 
To facilitate the computation of the two-point correlation functions we only consider two specific configurations of two galaxies in our numerical investigations. In one configuration the two galaxies are at the same redshift, i.e. $z_1 \equiv z_2$, and we study how the correlation functions change with the angular separation $\theta$, which is related to the comoving distance $r$ between the galaxies by the simple trigonometric relation $r \equiv \bar r_z \sqrt{2(1-\cos\theta)}$, where $\bar r_z \equiv \bar r_{z_1} \equiv \bar r_{z_2}$. In the other configuration the two galaxies lie on the same line of sight, i.e. $\boldsymbol{ \hat n}_1 \equiv \boldsymbol{ \hat n}_2$ $(\theta=0)$, but at different redshifts and we study how the correlation changes with the comoving separation $r = \bar r_{z_1}-\bar r_{z_2}$. In this case the redshift value $z_C$ of the middle point between the two galaxies is held fixed. These two configurations represent the two limiting cases of the general configurations of the two-point correlation function.

For numerical calculations we assume a flat $\Lambda$CDM universe with matter density $\Omega_m=0.3038$, baryon density $\Omega_b = 0.0462$, dark energy density $\Omega_\Lambda=0.65$, scalar amplitude $A_s=2.1 \times 10^{-9}$ at the pivot scale $k_0 = 0.05 \,\text{Mpc}^{-1}$, spectral index $n_s=0.96$, Hubble parameter $h=0.70$ and bias factor $b=2$ unless otherwise stated. 
Furthermore, we assume no magnification bias and the evolution bias $e_z=1.5$ at $z=1$, consistent with dark matter halos of bias $b=2$ in the Press-Schechter model \cite{Press:1973iz}.

\subsection{Contributions of the scalar perturbations}
\label{scalar contribution}

In this subsection we compute the scalar contributions to the two-point correlation function of the galaxy number density fluctuation $\langle \delta_g (z_1,\boldsymbol{\hat n}_1) \delta_g(z_2,\boldsymbol{\hat n}_2) \rangle$. In the conformal Newtonian gauge and with only scalar perturbations, the expression of the galaxy number density fluctuation is derived in sec.~\ref{GI formalism}:
\begin{equation}
\label{delta g num}
	\delta_g = (b \, \delta_m -e_z \, \delta z_v)   + \delta V\,, 
\end{equation}
where $\delta_m \equiv \delta_v$ and $\delta z_v=\delta z + \mathcal H \,v$, as explained in the last paragraph of sec.~\ref{sec: scalar}. The scalar contribution to the volume distortion is in turn given by 
\begin{equation}
\label{delta V num}
	\delta V =  3\, \delta z  + 2\, \frac{\delta r}{\bar r_z} - 2\,\mathcal K + H_z \frac{\partial}{\partial z} \delta r +V_{||} - 3\,\Psi\,,
\end{equation}
where the geometric distortions are expressed in terms of the gravitational potentials $\Psi$, the line-of-sight component of the peculiar velocities $V_{||} \equiv \hat n^i V_i$ and the coordinates lapses at the observer $\delta\eta_o$ and $\delta r_{o} \equiv \delta x_{||o} \equiv \hat n_i \delta x_{o}^i $ as
\begin{equation}
	\delta z = \mathcal H_o \delta\eta_o + \big[ V_{||} - \Psi \big]^z_o   
	  -  2 \int_0^{\bar r_z} d\bar r \, {\Psi}' \,, 
\end{equation}
\begin{equation}
	\delta r =  \delta r_{o} + \delta\eta_o - \frac{\delta z}{\mathcal{H}_z}
	 + 2 \int_0^{\bar r_z} d\bar r\,\Psi \,,
\end{equation}
\begin{equation}
		\mathcal K =  \frac{\delta r_{o}}{\bar r_z}  - V_{||o} 
		 +\int_0^{\bar r_z}d\bar r\, \bigg(\frac{\bar r_z - \bar r}{\bar r_z \bar r}\bigg)\hat\nabla^2 \Psi \,.
\end{equation}
As derived in appendix \ref{app solutions}, all the variables appearing in the above expressions can be expressed in terms of the (time-independent) curvature perturbation in the comoving gauge $\zeta (\boldsymbol{x})$, which is in turn related to the matter density contrast $\delta_+(\boldsymbol{x})$ at initial epoch. In Fourier space the latter is used to define the matter power spectrum $P_m(k)$ through $\langle \delta_+(\boldsymbol{k}_1)\delta_+(\boldsymbol{k}_2)\rangle \equiv (2\pi)^3 \delta_D(\boldsymbol{k}_1+\boldsymbol{k}_2)P_m(k_1)$, which allows to compute the two-point statistics by taking expectation values of the perturbations in conjunction with the corresponding growth factors.

\begin{figure}[h]
	\centering
		\includegraphics[scale = 0.65]{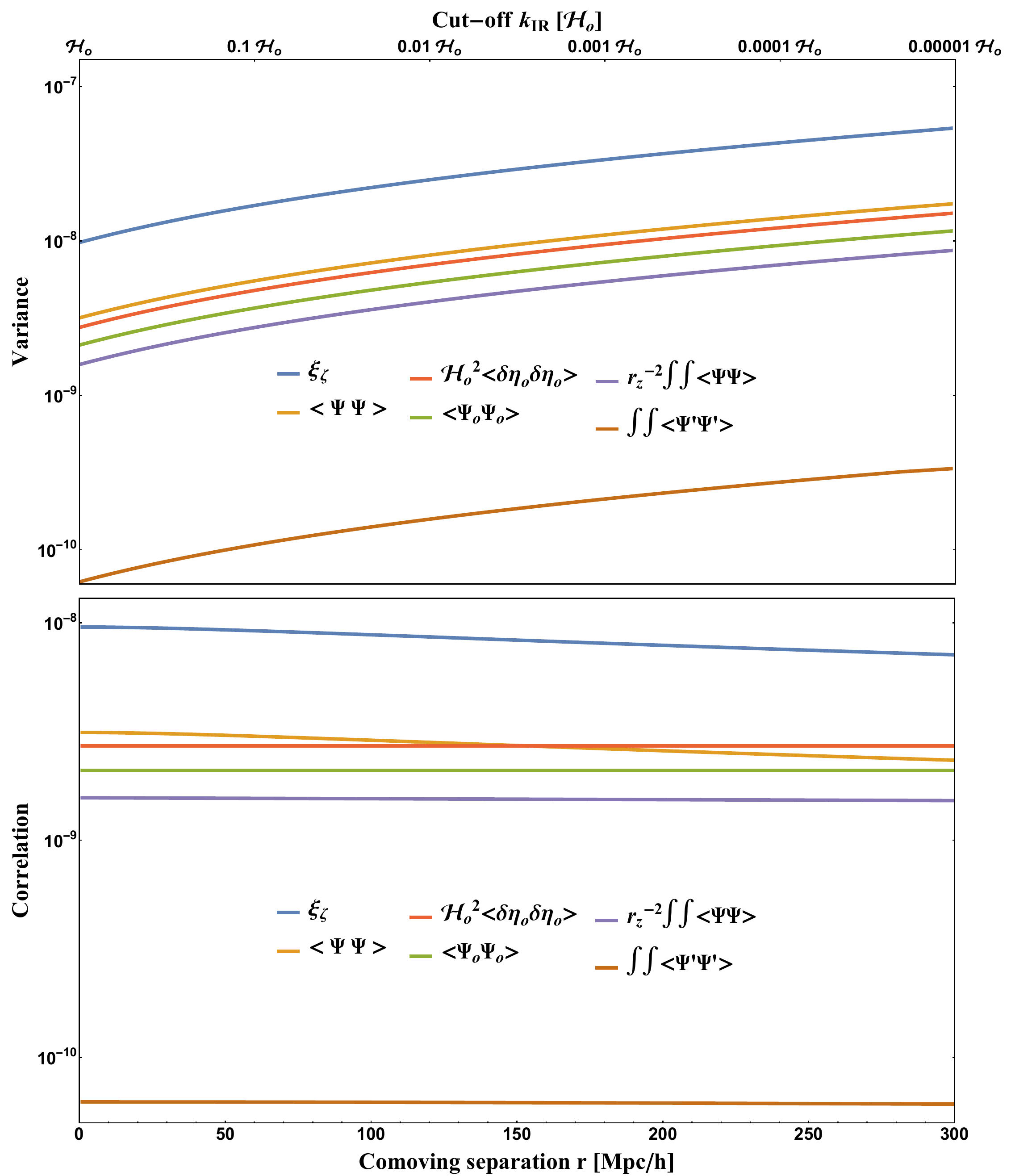}
		\caption{The top panel shows the dependence of the variance on the infrared cut-off for the curvature perturbation (blue), the potential at the source (orange), the time-lapse at the observer (red), the potential at the observer (green), the Shapiro time delay (purple) and the integrated Sachs-Wolf effect (brown). All these variances diverge when the IR cut-off $k_{\text{IR}}$ goes to zero. Though the individual terms diverge logarithmically, they add up to result in a finite contribution to the observed galaxy number density.
The bottom panel shows the auto-correlations of the same quantities as a function of the separation, when the infrared cut-off is set as $k_{\text{IR}}=\mathcal H_o$. We consider two galaxies at the same redshift $z_1=z_2=1$, so that the correlations are only functions of the comoving distance between two galaxies. Note that the correlation of the curvature perturbation does not depend on redshift, while those of the time-lapse and the potential at the observer are constant with the separation. The other correlations depend both on the redshift and the (spatial) separation. At redshift $z=1$ the value of the gravitational potential growth function is $D_\Psi = - 0.57$, where the negative value is due to the sign convention in appendix \ref{app solutions} and $D_\Psi=-0.6$ in the Einstein-de Sitter universe. The cut-off scale $k_{\text{IR}}=\mathcal H_o$ adopted here is rather arbitrary --- as shown in the upper panel, a larger-scale cut-off ($k_{\text{IR}}\ll \mathcal H_o$) would result in larger amplitudes of the correlation functions in the bottom panel.
		} 
	\label{fig cut-off div}
\end{figure}

\begin{figure}[h]
	\centering
		\includegraphics[width=\textwidth]{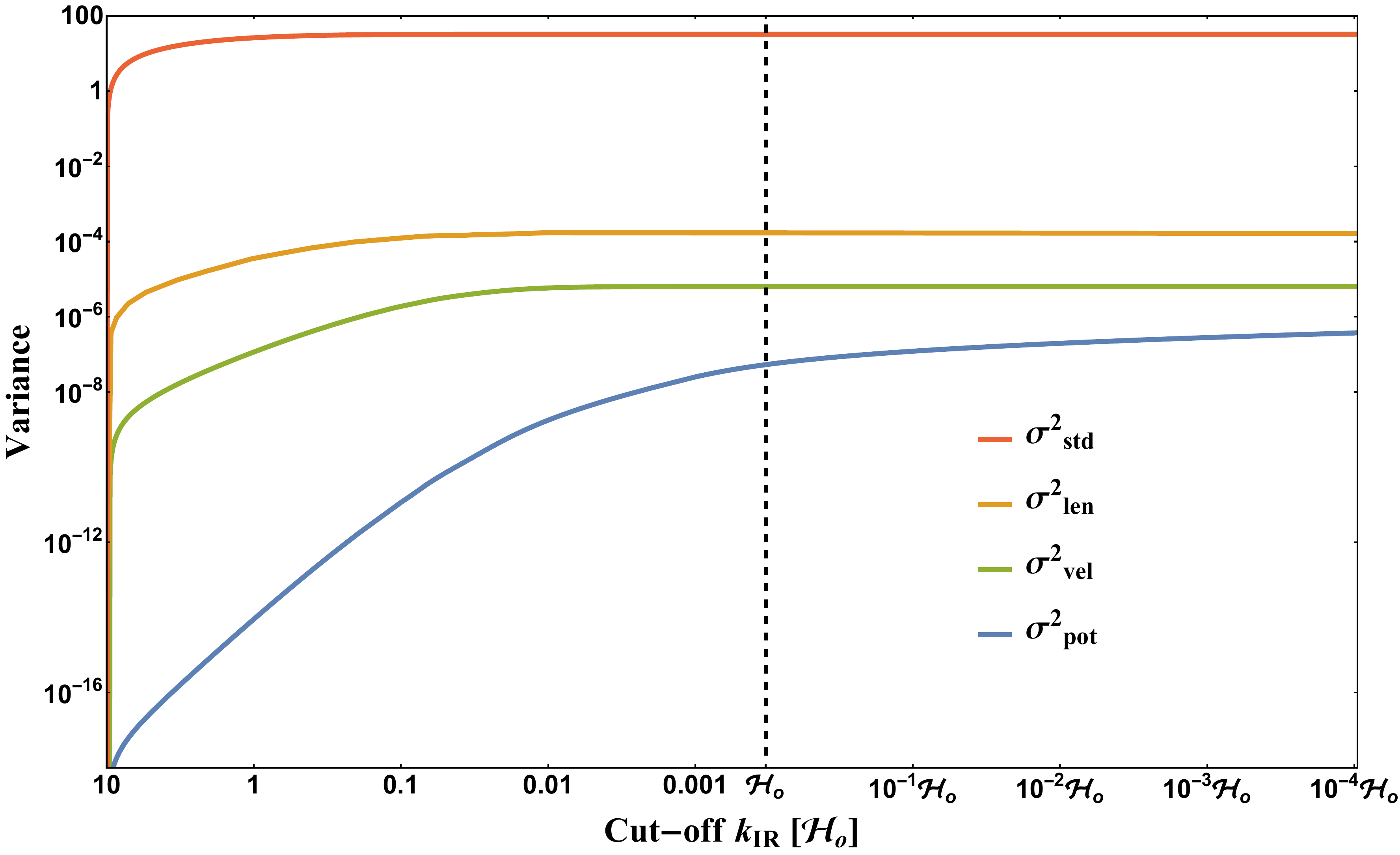}
		\caption{Individual contributions to the variance $\sigma^2=\langle \delta_g^2 \rangle$. The galaxy number density fluctuation is split as $\delta_g \equiv \delta_{\text{std}} + \delta_{\text{vel}}+ \delta_{\text{len}} + \delta_{\text{pot}}$, where the expression of each contribution is given in eq.~\eqref{var dg}. The figure shows the variances $\sigma^2_{i}=\langle \delta_i^2 \rangle$ of each contribution $\delta_i$ as functions of the IR cut-off of integration in Fourier space for galaxies at redshift $z=1$.  The vertical dashed line marks the horizon scale $k_{\text{IR}}=\mathcal H_o$, which represents our cut-off choice for the numerical computations of the correlation functions. The UV cut-off of integration is $k_{\text{UV}}\equiv 10 \, h/\mathrm{Mpc}$, so that the variances are vanishing when $k_{\text{IR}} = 10 \, h/\mathrm{Mpc}$.}
	\label{fig variances}
\end{figure}

As we showed in sec.~\ref{sec: scalar}, the monopole and dipole of long-mode perturbations do not affect the galaxy number density fluctuation, in agreement with the equivalence principle. Consequently, the total correlation function (including auto- and cross-correlations of all contributions) does not go to infinity when integrated over all $k$, because the divergences coming from the monopoles of different contributions cancel each other. 
Indeed, the correlations of quantities involving the potential $\Psi$ at the source, at the observer or integrated along the line of sight, as well as those involving the time lapse at the observer $\delta\eta_o$, diverge in the infrared, when $k$ is smaller than some value $k_{\text{IR}}$ close to zero. Only when these contributions are summed together the correlation function converges, because the effect of long-mode perturbations disappears, as also described in \cite{SG divergence}. 
The divergent behavior of the correlation function in the infrared, claimed in \cite{Bertacca:2012tp}, is due to the fact that terms evaluated at the observer position, such as $\Psi_o$ and $\delta\eta_o$, are usually neglected. The top panel of fig.~\ref{fig cut-off div} shows the dependence on the infrared cut-off for the variances of the terms discussed above, which blow up when $k_{\text{IR}}$ approaches zero.
The sum of all individually divergent contributions in the correlation is instead finite. 
As we show, these contributions turn out to be small compared to the density contribution, such that we set a sufficiently large yet arbitrary cut-off $k_{\text{IR}} \equiv \mathcal H_o$.
Indeed, as shown in fig.~\ref{fig variances}, the variance of the sum of the divergent contributions converges for $k_{\text{IR}} < \mathcal H_o$. In this plot the galaxy number density fluctuation is split as $\delta_g \equiv \delta_{\text{std}} + \delta_{\text{vel}} + \delta_{\text{len}} + \delta_{\text{pot}} \,$, where
\begin{equation}
\label{var dg}
\begin{split}
	\delta_{\text{std}} &=  b \, \delta_m -\frac{1}{\mathcal H_z}\partial_{||}V_{||} \,,
\qquad  
	\delta_{\text{vel}} =  h(z)  \big[ V_{||} \big]^z_o + 2 V_{||o}\, \,,
\qquad 
\delta_{\text{len}} =	- 2  \int_0^{\bar r_z} d\bar r  \bigg(\frac{\bar r_z - \bar r}{\bar r_z \bar r}\bigg)\hat\nabla^2 \Psi \,,
\\
\delta_{\text{pot}} &= \bigg[ h(z) - \frac{2}{\mathcal H_o \bar r_z} \bigg] \mathcal H_o \delta \eta_o +  e_z  \mathcal H_z v -  h(z) \big[\Psi\big]^z_o - \Psi   
	  + \frac{1}{\mathcal H_z} \Psi'
	+
	  \int_0^{\bar r_z} d\bar r\bigg[ \frac{4}{\bar r_z} \Psi  -  2\, h(z) \, {\Psi}' \bigg] \,,
	  \end{split}
\end{equation}
and we have defined the function of redshift $h(z) \equiv 3-e_z - \mathcal H_z' / \mathcal H_z^2 -  2 / (\bar r_z  \mathcal{H}_z)$. All the perturbations with divergent individual correlation, or variance (see top panel of fig.~\ref{fig cut-off div}), are contained in $\delta_{\text{pot}}$. When we compute the corresponding variance $\sigma^2_{\text{pot}}=\langle \delta_{\text{pot}}^2 \rangle$ these contributions are summed together before taking the ensemble average, leading to a convergent result. 
This might not be perfectly represented in fig.~\ref{fig variances}, due to numerical residuals in the evaluation of the integrands in the variance expression. Indeed, to compute the variance we split the time and space dependence in the perturbations, using the growth functions defined in appendix \ref{app solutions}. Therefore, the variance is given by time-dependent factors multiplied by integrals over Fourier modes of the time-independent part of the perturbations. However, as shown in sec.~\ref{EP}, the time-dependent factor that multiplies the divergent integrations is exactly zero.
As a conclusion, the theoretical prediction for the correlation function of the galaxy number density is gauge-invariant and finite, provided that we take into account all terms in the relativistic derivation. In practice, the standard way of computing the variance by using $\delta_{\text{std}}$ alone is accurate at the 1\% level, and the dominant correction originates from the lensing convergence $\delta_{\text{len}}$. For the computation of the gravitational potential contribution $\delta_{\text{pot}}$ our numerical calculations demonstrate that one can safely impose an IR cut-off scale, as long as $k_{\text{IR}} \lesssim  \mathcal H_o$.

\subsubsection{Matter density fluctuation}

We now want to study the two-point correlation functions of the various quantities in eqs.~\eqref{delta g num} and \eqref{delta V num}. 
Before proceeding it is convenient to split the time and space dependences in the perturbations as in sec.~\ref{sec: scalar} and appendix \ref{app solutions}. In this way, the dependence on the redshifts $z_1$, $z_2$ in the correlation functions can be factorized through the growth functions $D$, $D_\Psi$, $D_V$ of the matter density contrast, the gravitational potential and the peculiar velocity, respectively. 
The auto-correlation function of the matter density contrast is then given by
\begin{equation}\label{densdens}
	\langle \delta_m (z_1, \boldsymbol{\hat n}_1) \delta_m(z_2, \boldsymbol{\hat n}_2) \rangle = D(z_1) D(z_2) \int_{k_{\text{IR}}}^{k_{\text{UV}}} \frac{dk}{2\pi^2} k^2 P_m(k) j_0 (kr) = D(z_1) D(z_2) \, \xi_m (r) 	  
	\,,
\end{equation}
where $r$ is the length of the spatial separation $\boldsymbol{r}=\bar r_{z_1} \boldsymbol{ \hat n}_1 - \bar r_{z_2} \boldsymbol{ \hat n}_2 $ between the two galaxies, $P_m(k)$ is the matter power spectrum computed using \texttt{CAMB}, $j_0(x)$ is the spherical Bessel function and in the last equality we have introduced the matter correlation function $\xi_m(r) = \langle \delta_+(\boldsymbol{x}) \delta_+(\boldsymbol{x}+\boldsymbol{r})\rangle$ at initial epoch. 
In the numerical evaluation we always set the lower and upper cut-offs in the integration as $k_{\text{IR}}\equiv\mathcal H_o = 100/c \,\,  \mathrm{km/s} \,\, h/\mathrm{Mpc} = 3.3 \times 10^{-4} \, h/\mathrm{Mpc}$ and $k_{\text{UV}}\equiv 10 \, h/\mathrm{Mpc}$, where $c$ is the speed of light and $h$ is the reduced Hubble constant.\footnote{The convergence of the correlations in the ultraviolet regime occurs at around $ k \approx 1 \, h/\mathrm{Mpc}$, but choosing a bigger value results in a more accurate numerical evaluation of the integrals.} 

\begin{figure}[h]
	\centering
		\includegraphics[width=\textwidth]{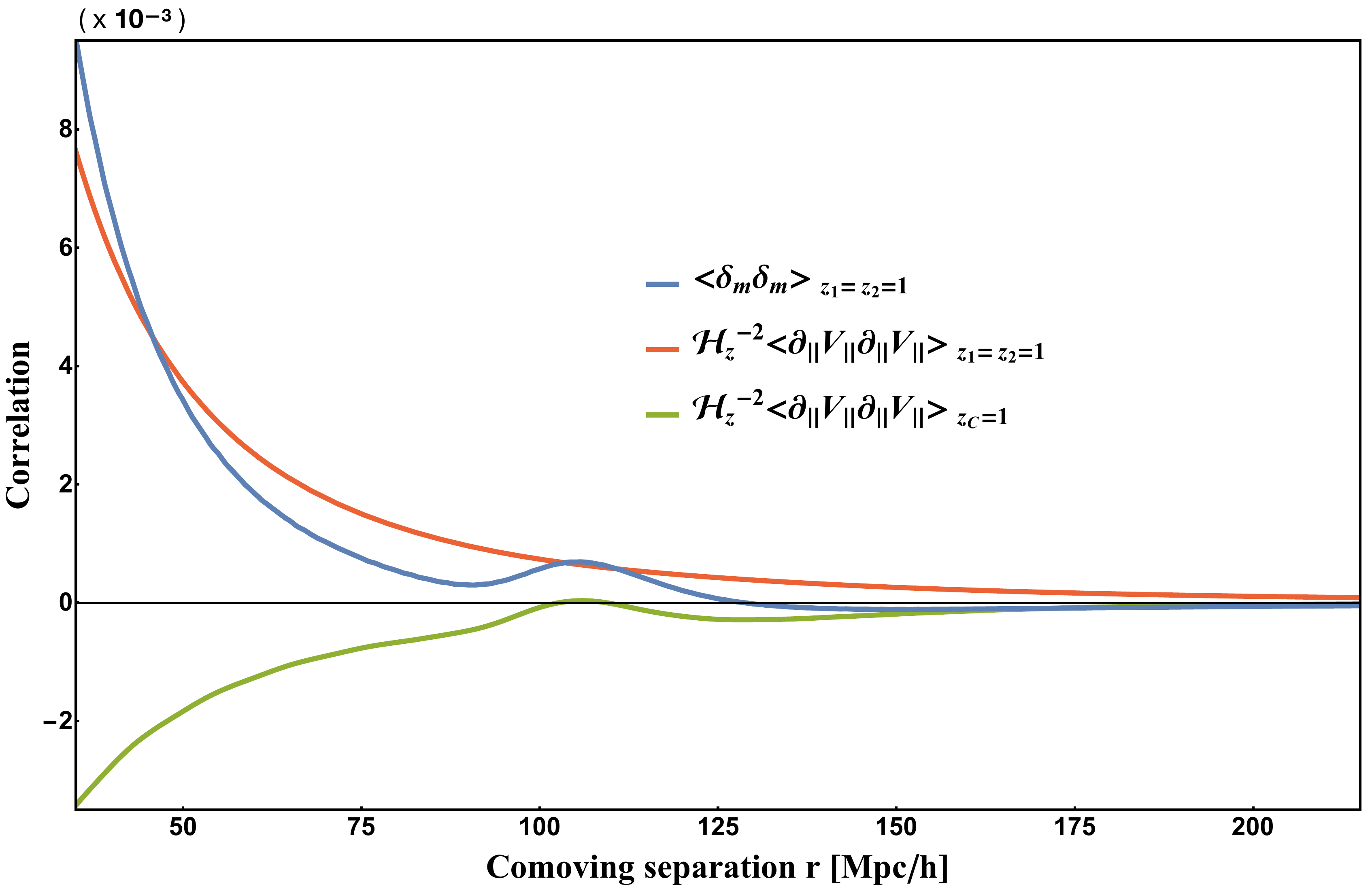}
		\caption{The correlation function of the density fluctuations for two galaxies at the same redshift $z_1=z_2=1$ is shown in blue. The correlation function of the Kaiser effects for two galaxies at the same redshift $z_1=z_2=1$ is shown in red, while that of two galaxies lying on the same line-of-sight with middle point at redshift $z_C=1$ is shown in green. }
	\label{fig dens kaiser}
\end{figure}

The behavior of the density auto-correlation given in eq.~\eqref{densdens} is shown by the blue curve in fig.~\ref{fig dens kaiser} as a function of the comoving separation $r$ between two galaxies at the same redshift $z_1=z_2=1$. The local maximum at around 110 Mpc/$h$ is there a well-known feature of the matter correlation function due to baryon acoustic oscillations (BAO). At around $r=130$ Mpc/$h$ the correlation is zero, because at this scale there is no deviation from a uniform distribution of galaxies (galaxies do not cluster). At larger scales, the correlation becomes negative, as galaxies tend to avoid each other. Obviously, as the separation increases further the anti-correlation between the density fields at the two end points decreases and reaches zero asymptotically. The other functions in the plot (red and green curves) are the two-point correlations of redshift-space distortions (for two different configurations) and we will discuss them in sec.~\ref{rsd}. The density contrast and the redshift-space distortion are the dominant contributions to the two-point correlation function of the galaxy number density and are devoid of any divergence both at IR and UV scales.

\subsubsection{Redshift and radial distortions}
We now want to study the correlation functions of the redshift distortion $\delta z$ and the radial distortion $\delta r$. Although these are gauge-dependent quantities, studying their correlation function is essential to understand how different relativistic effects contribute to the total galaxy correlation function. The analytic expressions are given by the sum of the auto- and cross-correlations of different quantities:
\begin{equation}
	\begin{split}
		\langle \delta z (z_1, \boldsymbol{\hat n}_1) \delta z (z_2, \boldsymbol{\hat n}_2) \rangle = \mathcal H_o^2 \langle \delta\eta_o \delta\eta_o \rangle + \langle V_{||1}  V_{||2}  \rangle + \langle V_{||o} (\boldsymbol{\hat n}_1) V_{||o} (\boldsymbol{\hat n}_2) \rangle    + \langle \Psi_1 \Psi_2 \rangle  + \langle \Psi_{o} \Psi_{o} \rangle &
	\\
	 + 4 \,  \int_0^{\bar r_{z_1}} d\bar r_1 \int_0^{\bar r_{z_2}} d\bar r_2 \, \langle{\Psi}'(\bar\eta_o - \bar r_1,\bar r_1 \, \boldsymbol{\hat n}_1 )  {\Psi}'(\bar\eta_o - \bar r_2,\bar r_2 \, \boldsymbol{\hat n}_2 ) \rangle
	 + \text{cross-correlations}  &, 
	\end{split}
\end{equation}
where we have introduced the notation $X_1 \equiv X(z_1, \boldsymbol{\hat n}_1)$ for any perturbation $X$,

\begin{equation}
	\begin{split}
		 \frac{\langle\delta r (z_1, \boldsymbol{\hat n}_1) \delta r (z_2, \boldsymbol{\hat n}_2) \rangle}{\bar r_{z_1}\bar r_{z_2}}  &=   \frac{\langle \delta r_{o}(\boldsymbol{\hat n}_1) \delta r_{o} (\boldsymbol{\hat n}_2) \rangle}{\bar r_{z_1}\bar r_{z_2}}   + \frac{\langle\delta\eta_o \delta\eta_o \rangle}{\bar r_{z_1}\bar r_{z_2}} +    \frac{\langle \delta z_1 \delta z_2 \rangle}{\bar r_{z_1}\mathcal{H}_{z_1} \bar r_{z_2}\mathcal{H}_{z_2}}   
	\\  
	& \quad + \frac{4}{\bar r_{z_1} \bar r_{z_2}}\,  \int_0^{\bar r_{z_1}} d\bar r_1 \int_0^{\bar r_{z_2}} d\bar r_2 \,\langle{\Psi}_1  {\Psi}_2 \rangle
	+ \text{cross-correlations} . 
	\end{split}
\end{equation}
Therefore, in order to understand which are the leading contributions to the correlations of $\delta z$ and $\delta r$, we have to compute the auto-correlation functions of the time-lapse at the observer $\delta\eta_o$, the peculiar velocities $V_{||}$, $V_{||o}$, the local potentials $\Psi$, $\Psi_o$, the integrated Sachs-Wolfe (ISW) effect $\int d\bar r\, \Psi'$, the spatial shift at the observer $\delta r_{o}$ and the Shapiro time-delay effect $\int d \bar r/\bar r_z\,\Psi$. 

First of all, the auto-correlations of the potential and the time-lapse at the observer are given by  
\begin{eqnarray}\label{tl}
	\langle \Psi_o \Psi_o \rangle &=& C^2 D_{\Psi_o}^2
 \int_{k_{\text{IR}}}^{k_{\text{UV}}} \frac{dk}{2\,\pi^2}\frac{1}{k^2} P_m(k) \,,
	\\
	\mathcal H_o^2 \langle \delta\eta_o \delta\eta_o \rangle &=& \mathcal H_o^2 \, C^2 D_{Vo}^2 \int_{k_\text{IR}}^{k_\text{UV}}\frac{dk}{2\pi^2} \frac{1}{k^2} P_m(k) \,,
\end{eqnarray}
where $C\equiv - {\mathcal H}^2 D f \,\Sigma$ is a constant and it becomes
$-\frac52{\cal H}_o^2\Omega_m$ in the matter-dominated universe (see appendix \ref{app solutions}). Note that, while the growth function $D_\Psi$ is dimensionless, $D_V$ has the same dimension as $\mathcal H^{-1}_o$. These correlations are then dimensionless and both independent of separation, adding up to a constant contribution when the IR cut-off is imposed, but they are divergent when the integration is performed from $k=0$, as shown in fig.~\ref{fig cut-off div}. As explained above, the divergence due to these quantities at the observer cancel the divergence due to the potential at the source and integrated along the line of sight. It is therefore important to consider $\Psi_o$ and $\delta\eta_o$ in the expression of the galaxy number density, also from the numerical point view. 

Let us now consider the correlations of the gravitational potential at the source (local) and integrated along the line of sight (non-local).
The auto-correlation of the gravitational potential at the source is given by 
\begin{equation}
	\langle \Psi (z_1, \boldsymbol{\hat n}_1) \Psi (z_2, \boldsymbol{\hat n}_2) \rangle = C^2 D_\Psi(z_1) D_\Psi(z_2) \int_{k_{\text{IR}}}^{k_{\text{UV}}} \frac{dk}{2\,\pi^2}\frac{1}{k^2} P_m(k) j_0(kr) 
	= D_\Psi(z_1) D_\Psi(z_2)\, \xi_\zeta (r) \,,
\end{equation}
where $\xi_\zeta (r)=\langle \zeta(\boldsymbol{x})\zeta(\boldsymbol{x}+\boldsymbol{r})\rangle$ is the correlation function of the curvature perturbation.
The non-local terms are the ISW and the Shapiro time-delay. Their auto-correlations are respectively given by 
\begin{equation}
\begin{split}
	\int_0^{\bar r_{z_1}}& d \bar r_1  \int_0^{\bar r_{z_2}} d\bar r_2 \,\langle {\Psi}'(\bar\eta_1,\bar r_1 \, \boldsymbol{\hat n}_1 )  {\Psi}'(\bar\eta_2,\bar r_2 \, \boldsymbol{\hat n}_2 ) \rangle 
	\\
	= & \int_0^{\bar r_{z_1}} d\bar r_1 \int_0^{\bar r_{z_2}} d\bar r_2 \, D'_{\Psi}(\bar \eta_o - \bar r_1) D'_{\Psi}(\bar \eta_o - \bar r_2) \int_{k_\text{IR}}^{k_\text{UV}}\frac{dk}{2\pi^2}  \frac{1}{k^2}P_m(k) j_0 (k \vert \bar r_1 \boldsymbol{\hat n_1} - \bar r_2 \boldsymbol{\hat n_2} \vert ) \,,
\end{split}
\end{equation}
\begin{equation}
\begin{split}\label{std}
\int_0^{\bar r_{z_1}} \frac{d\bar r_1}{\bar r_{z_1}}  \int_0^{\bar r_{z_2}} \frac{d\bar r_2}{ \bar r_{z_2}}  \,\langle  {\Psi} &(\bar\eta_1,\bar r_1 \, \boldsymbol{\hat n}_1 )  {\Psi}(\bar\eta_2,\bar r_2 \, \boldsymbol{\hat n}_2 ) \rangle 
\\
&= \int_0^{\bar r_{z_1}} \frac{d\bar r_1}{\bar r_{z_1}} \int_0^{\bar r_{z_2}} \frac{d\bar r_2}{ \bar r_{z_2}}  \,  D_\Psi(\bar \eta_o - \bar r_1) D_\Psi(\bar \eta_o - \bar r_2) \xi_\zeta (\vert \bar r_1 \boldsymbol{\hat n_1} - \bar r_2 \boldsymbol{\hat n_2} \vert) \,.
\end{split}
\end{equation}
Fig.~\ref{fig cut-off div} shows the variances $( r \rightarrow 0)$ of these contributions as a function of the IR cut-off (top panel) as well as the correlations as functions of $r$, when $k_{\text{IR}}\equiv \mathcal H_o$, for two galaxies at the same redshift $z_1=z_2=1$ (bottom panel). One should notice that, while the correlations of the potentials and the time-lapses at the observer are exactly constant, the correlations of ISW and Shapiro time-delay effects vary as a function of scale, though the change is too small to be visible in the range of separations considered, except the correlation of the potential at the sources. Nevertheless, all these correlations are from five to seven orders of magnitude smaller the matter density correlation (compare with fig.~\ref{fig dens kaiser}).

\begin{figure}[h]
	\centering
		\includegraphics[width=\textwidth]{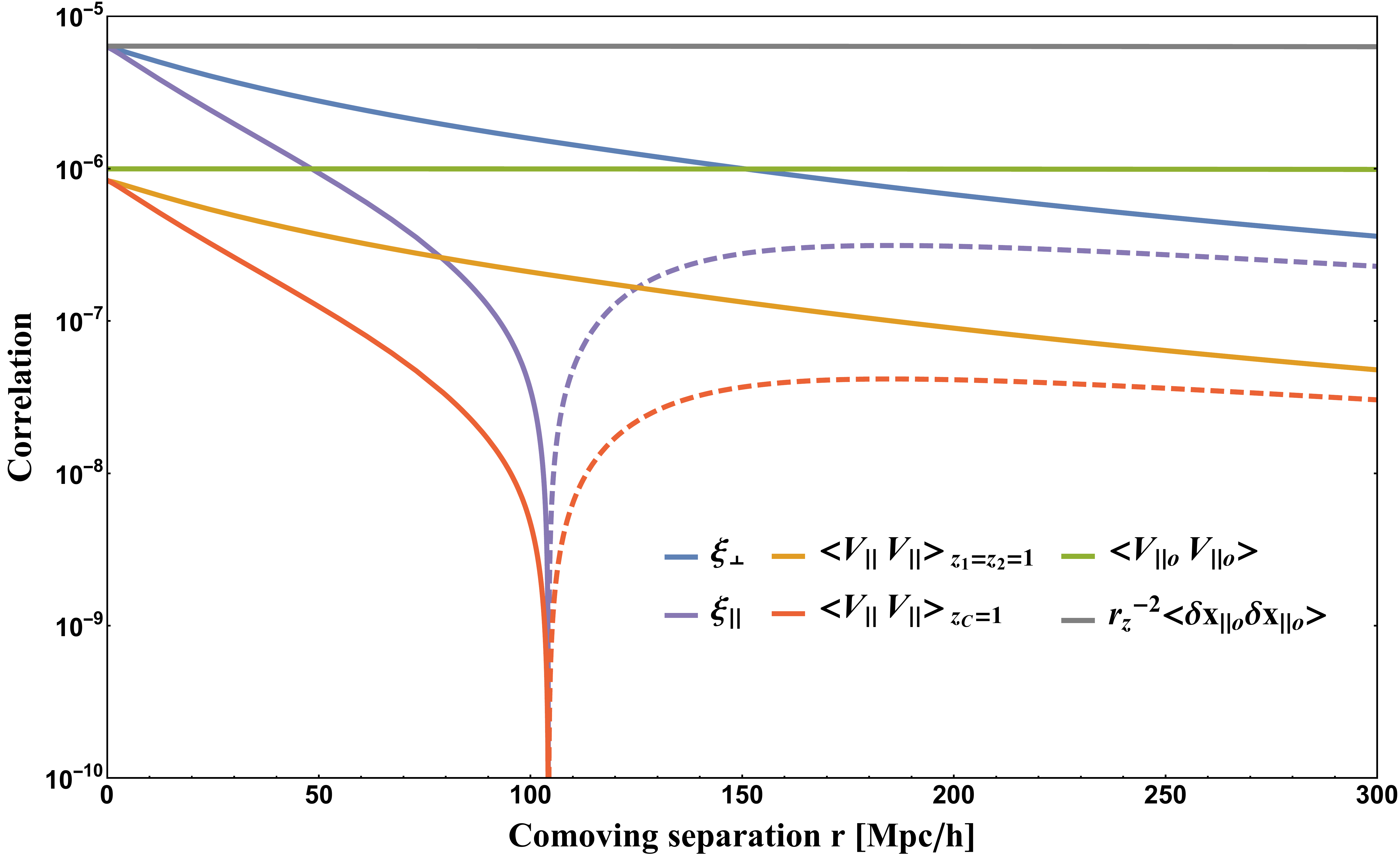}
		\caption{The velocity correlation function is decomposed into $\xi_{||}$ and $\xi_\perp$, respectively parallel and perpendicular to the separation between the two points under consideration. The correlation function of the velocities of two galaxies at the same redshift $z_1=z_2=1$ is shown in orange, while that of two galaxies lying on the same line-of-sight with middle point at redshift $z_C=1$ is shown in red.  Dashed lines represent negative values. The correlation functions of the velocities and the spatial shifts at the observer point are shown in green and gray respectively. These are only functions of the angle between the two lines of sight and the relation of this variable with the separation $r$ depends on the redshift considered. 
Hence they vary as a function of $r$, but very little over the range in this plot. 
The non-trivial factors with which the correlation functions $\langle V_{||}V_{||} \rangle$ and $\langle V_{||o}V_{||o} \rangle$ appear in the expressions are the function of redshift $\mathcal H_z'/\mathcal H_z^2$ or $1/(\mathcal H_z \bar r_z)$, and their values at $z=1$ are respectively $-0.16$ and $1.47$.
The value of $\bar r_z$ in $\bar r_z^2 \langle \delta r_o \delta r_o \rangle$ is that at redshift $z=1$, for consistency with the other functions in the plot. Note that at $r=0$ the amplitude of $\bar r_z^2 \langle \delta r_o \delta r_o \rangle$ is not the same as $\xi_{||}$ and $\xi_{\perp}$, as might appear from the plot.
		The cut-off choices are $k_{\mathrm{IR}}=\mathcal H_o$ and $k_{\mathrm{UV}}=10$ $h$/Mpc.}
	\label{fig V}
\end{figure}

The remaining contributions to compute are those of the velocities and the spatial shifts at the observer, which are finite even when the integration is performed from $k_{\text{IR}}=0$. 
By applying the velocity solution $V_{||}(\eta,\boldsymbol{x})=D_{V}(\eta)\,\partial_{||}\zeta(\boldsymbol{x})$ (see appendix \ref{app solutions}), the correlation function of the two line-of-sight velocities can be written as (see appendix \ref{app derivations} for the derivation)
\begin{equation}\label{v corr}
	\langle V_{||} (z_1, \hat n_1) V_{||}(z_2, \hat n_2) \rangle = \bigg( \frac{C}{\mathcal H_o} \bigg)^2 D_V(z_1) D_V(z_2) \big\lbrace \hat{\mathcal P}_{||} \xi_{||}(r) + \hat{\mathcal P}_{\perp} \xi_{\perp}(r) \big\rbrace  \,,
\end{equation}
where, by defining $\hat{\mathcal P}_{||} \equiv \hat n^i_1 \hat n^j_2\, \hat r_i \hat r_j$ and $\hat{\mathcal P}_{\perp} \equiv \hat n^i_1 \hat n^j_2\, (\delta_{ij}- \hat r_i \hat r_j)$ as in \cite{SG correlation}, we decomposed the velocity correlation function into the parallel and perpendicular components with respect to the separation $\boldsymbol{r}\,$: 
\begin{equation}
	\xi_{||}(r) \equiv -\mathcal H_o^2 \int_{k_\text{IR}}^{k_\text{UV}}\frac{dk}{2\pi^2} P_m(k) \frac{j_0' (kr)}{kr}\,, \qquad\qquad \xi_{\perp}(r) \equiv -\mathcal H_o^2 \int_{k_\text{IR}}^{k_\text{UV}}\frac{dk}{2\pi^2} P_m(k) j_0'' (kr) \,,
\end{equation}
with $j_0'(x)=\partial_x j_0(x)$ and $j_0''(x)=\partial_x^2 j_0(x)$. Note that our expression is derived without assuming the distant-observer approximation and it is valid for any two lines of sight.
Fig.~\ref{fig V} shows the behavior of $\xi_{||}$ and $\xi_{\perp}$ with respect to $r$, as well as the correlations of velocities at the sources (for the two configurations (i) $z_1=z_2=1$ and (ii) $\boldsymbol{\hat n}_1=\boldsymbol{\hat n}_2$ with middle point between the two galaxies at redshift $z_C=1$) and at the observer. The latter is only a function of the angle $\theta = \cos^{-1}(\boldsymbol{\hat n_1} \cdot \boldsymbol{\hat n_2})$ between the two lines of sight,
\begin{equation}
	\langle V_{||o} (\boldsymbol{\hat n}_1) V_{||o} (\boldsymbol{\hat n}_2) \rangle  = \boldsymbol{\hat n_1} \cdot \boldsymbol{\hat n_2}\,\, C^2 D_{Vo}^2  \int_{k_\text{IR}}^{k_\text{UV}}\frac{dk}{2\pi^2}\, \frac{1}{3}P_m(k)  \,,
\end{equation}
and therefore it varies very little in the range of $r$ considered, despite not being constant. Note, however, that the relation between the angle $\theta$ and the separation $r$ depends on the redshift (the higher is the redshift, the larger is the separation associated to a given angle at the observer) and in fig.~\ref{fig V} we take $z=1$.

The correlation of the spatial shift at the observer is also only a function of the angle between the two lines of sight, given by
\begin{equation}
	\langle \delta r_{o}(\boldsymbol{\hat n_1})\delta r_{o}(\boldsymbol{\hat n_2}) \rangle = \boldsymbol{\hat n_1} \cdot \boldsymbol{\hat n_2} \,\, \bigg( C \int_0^{\bar \eta_o} d\bar \eta \, D_V(\bar \eta) \bigg)^2 \int_{k_\text{IR}}^{k_\text{UV}}\frac{dk}{2\pi^2}\, \frac{1}{3}P_m(k)   \,.
\end{equation}
In fig.~\ref{fig V} one can see that, when considering two galaxies at redshift $z_1=z_2=1$, the correlation of this effect is higher than that of the velocities at the observer by almost one order of magnitude. Note, additionally, that it can be much higher than the correlation of velocities at the sources if a large separation is considered. Indeed, at $r=300$ Mpc/$h$, the difference is given by a factor of almost 20.

The spatial shift and the velocity at the observer position are typically ignored in literature, but their contributions are larger than the velocity correlation of two source galaxies. However, note that the spatial shift at the observer cancels at the linear order with the same term in the lensing convergence, such that ignoring this contribution does not cause any systematic error. On the other hand, the contribution of the velocity at the observer (green line in fig.~\ref{fig V}) must be kept in the calculations, and it is larger than the velocity correlation (orange line in fig.~\ref{fig V}).

\begin{figure}[h]
	\centering
		\includegraphics[width=\textwidth]{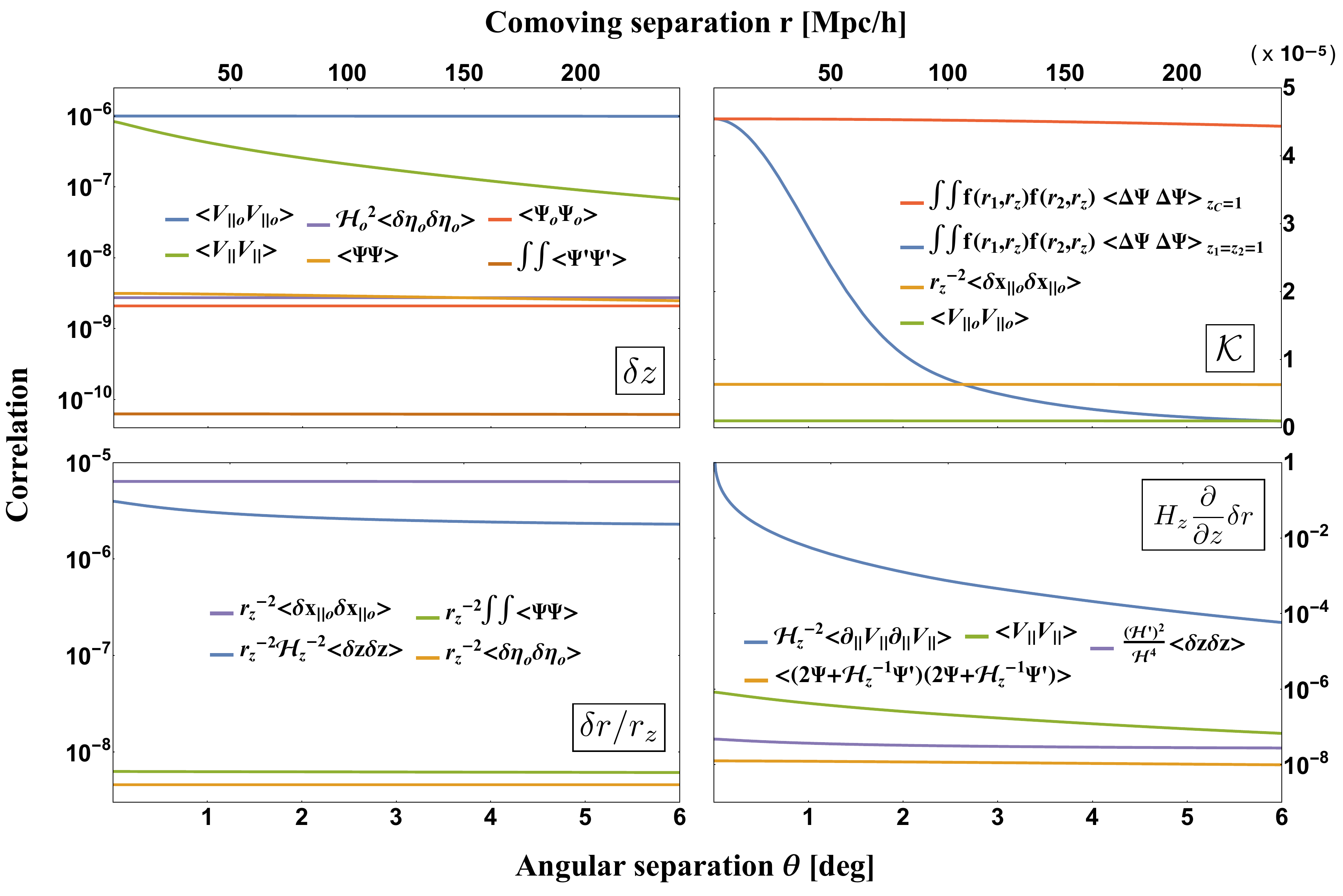}
		\caption{Top left panel: auto-correlations of various contributions to $\langle \delta z \delta z \rangle$ as functions of the separation between two galaxies at the same redshift $z_1=z_2=1$. 
		Bottom left panel: auto-correlations of various contributions to $\langle \delta r \delta r \rangle /\bar r_{z}^2$ as functions of the separation between two galaxies at the same redshift $z_1=z_2=1$. 
		Top right panel: The correlation function of the lensing contribution $\int_0^{\bar r_z}d\bar r\,f(\bar r,\bar r_z) \Delta \Psi $, where $f(\bar r,\bar r_z)= (\bar r_z - \bar r)\frac{\bar r}{\bar r_z}$, for two galaxies at the same redshift $z_1=z_2=1$ is shown in blue, while that of two galaxies lying on the same line-of-sight with middle point at redshift $z_C=1$ is shown in red (note that the latter is only a function of the separation $r$, as in this configuration $\theta=0$). Auto-correlations of other contributions to the gravitational lensing convergence $\mathcal K$ are shown in orange and green. Bottom right panel: auto-correlations of various contributions to $H_z^2\langle \partial_z \delta r \, \partial_z \delta r \rangle$ as functions of the separation between two galaxies at the same redshift $z_1=z_2=1$. The dominant contribution is given by the Kaiser effect.}
           	\label{fig dist}
\end{figure}

We now have the main ingredients to analyze the correlations of the redshift and the radial distortions. Clearly, one has to compute also cross-correlations among all terms considered so far. 
Nevertheless, by looking at the auto-correlations of individual contributions to $\delta z$ and $\delta r$ one can obtain a clear intuition of the importance of each effect in the correlations $\langle \delta z_1 \delta z_2 \rangle$ and $\langle \delta r_1 \delta r_2 \rangle / \bar r_{z_1} / \bar r_{z_2}$.
The top left panel of fig.~\ref{fig dist} shows the auto-correlations of all contributions to $\delta z$ for two galaxies at the same redshift $z_1=z_2=1$. We immediately see that the correlation of the redshift distortion $\langle \delta z_1 \delta z_2 \rangle$ is dominated by the Doppler effect of peculiar velocities, including that by the observer velocity, as expected. Indeed, the correlations of local potentials and the time-lapse at the observer are about 2 orders of magnitude smaller than the correlation of velocities at the observer, representing the leading contribution. Compared to the latter the correlation of the ISW effect is even 4 orders of magnitude smaller.

Analogously, the bottom left panel of fig.~\ref{fig dist} shows the auto-correlations of all contributions to $\delta r$ for the same configuration ($z_1=z_2=1$). Evidently, the strongest contribution to the correlation of the radial distortions comes from the spatial shift at the observer. However, we again emphasize that the latter is absent in the total expression of the galaxy number density, because the same term appears in the gravitational lensing convergence $\mathcal K$ with opposite sign. The correlation of redshift distortions $\langle \delta z_1 \delta z_2 \rangle /\bar r_{z}^2 /\mathcal H_{z}^2$ is smaller than that of the spatial shift at the observer but it contributes with the same order of magnitude ($10^{-6}-10^{-5}$) to $\langle \delta r_1 \delta r_2 \rangle / \bar r_{z}^2$. So, the correlation of radial distortions, like that of redshift distortions, is dominated by the effect of velocities.

Note, finally, that both the redshift and the radial distortions are not directly observables, they are affected by the long-mode perturbations (see sec.~\ref{EP}) and their correlations are divergent in the infrared. In other words, the sum of the correlations in eqs.~\eqref{tl}$-$\eqref{std} and their cross-correlations diverges if the IR cut-off is removed. Such divergence is eliminated when the remaining contributions to the galaxy number density fluctuations are taken into account in the correlation.

We want to emphasize that all the individual components such as $\delta z$, $\delta r$ and so on are gauge-invariant in the Newtonian gauge, but they diverge in the infrared: gauge-invariance is not a sufficient condition for observable quantities. Furthermore, this decomposition of the observable galaxy number density depends on \textit{our gauge choice}, in the sense that while the expressions for $\delta z$ in the Newtonian gauge or comoving gauge, for instance, are gauge-invariant, their values are different.

\subsubsection{Lensing convergence}

The next effect to consider in the expression of the galaxy number density is the gravitational lensing convergence.
To obtain the correlation function of the lensing convergence it is convenient to first express it as follows:
\begin{equation}
	\mathcal K = \frac{\delta r_{o}}{\bar r_z}  - V_{||o}  +  2\,\Psi_o - \Psi +  \int_0^{\bar r_z}d\bar r\,\bigg[ - 2\frac{\bar r}{\bar r_z} \Psi' -  \bigg(1 - \frac{\bar r}{\bar r_z}  \bigg)\bar r \, \Psi'' 
		 + (\bar r_z - \bar r)\frac{\bar r}{\bar r_z}\Delta \Psi \bigg] \,,
\end{equation}
where we have used the relation between the angular Laplacian $\hat \nabla^2$ and the 3D Laplacian $\Delta$:
\begin{equation}
	\Delta = \frac{1}{\bar r^2} \hat \nabla^2 + \frac{2}{\bar r}  \frac{\partial}{\partial \bar r} +  \frac{\partial^2}{\partial \bar r^2} \,.
\end{equation}
In this way we can use the Poisson equation $(\Delta \Psi = \frac{3}{2}\mathcal H_o^2 \Omega_m \delta / a)$ and write the correlation function as
\begin{equation}
\begin{split}
	\langle \mathcal K(z_1, \boldsymbol{\hat n}_1) \mathcal K(z_2, \boldsymbol{\hat n}_2) \rangle &= \frac{\langle \delta r_{o}\delta r_{o} \rangle}{\bar r_{z_1}\bar r_{z_2}} + \langle V_{||o} (\boldsymbol{\hat n}_1) V_{||o} (\boldsymbol{\hat n}_2) \rangle
	 + 4\, \langle \Psi_{o} \Psi_{o} \rangle + \langle \Psi_1 \Psi_2 \rangle 
	\\
	&\quad + 4 \,  \int_0^{\bar r_{z_1}}d\bar r_1 \int_0^{\bar r_{z_2}}d\bar r_2\, \frac{\bar r_1 \bar r_2}{\bar r_{z_2} \bar r_{z_1}}  \langle \Psi' (\bar\eta_1,\bar r_1 \, \boldsymbol{\hat n}_1)\Psi' (\bar\eta_2,\bar r_2 \, \boldsymbol{\hat n}_2) \rangle 
	\\
	&\quad +  \int_0^{\bar r_{z_1}} d\bar r_1\, \frac{(\bar r_{z_1}-\bar r_1)\bar r_1}{\bar r_{z_1}}  \int_0^{\bar r_{z_2}} d\bar r_2\, \frac{(\bar r_{z_2}-\bar r_2)\bar r_2}{\bar r_{z_2}} \langle \Psi''(\bar\eta_1,\bar r_1 \, \boldsymbol{\hat n}_1) \Psi'' (\bar\eta_2,\bar r_2 \, \boldsymbol{\hat n}_2)  \rangle
	\\
	&\quad + \frac{9}{4}\mathcal H_o^4 \Omega_m^2\int_0^{\bar r_{z_1}}d\bar r_1 \, g(\bar r_1) \int_0^{\bar r_{z_2}}d\bar r_2 \, g(\bar r_2) \, \xi_m(\vert  \bar r_1 \boldsymbol{\hat n_1} - \bar r_2 \boldsymbol{\hat n_2} \vert )
	\\
	& \quad + \text{cross-correlations} \,,
\end{split}
\end{equation}
where we have defined $g(\bar r_i)\equiv \frac{(\bar r_{z_i}-\bar r_i)\bar r_i}{\bar r_{z_i}}\frac{D(\bar\eta_o - \bar r_i)}{a(\bar\eta_o - \bar r_i)}$. 
While the first two terms have been already discussed, the remaining ones constitute together the convergence of light rays $\int_0^{\bar r_z}d\bar r\, (\frac{\bar r_z - \bar r}{\bar r_z \bar r})\hat\nabla^2 \Psi$. These terms do not lead to a divergence in the correlation when $k_{\text{IR}}\rightarrow 0$. 
Indeed, as confirmed by our analysis in sec.~\ref{sec: scalar} (see in particular eq.~\eqref{scalar mon dip}), the lensing convergence does not contain the monopole of the long-mode gravitational potential. This is due to the fact that spatial derivatives of the potential are involved in the expression of $\mathcal K$, which gives zero when applied to the monopole.

The top right panel in fig.~\ref{fig dist} shows the auto-correlations of the three contributions to $\mathcal K$: the spatial shift at the observer, the velocity at the observer and the non-local convergence. Again, two galaxies at redshift $z_1=z_2=1$ are considered and the correlations are therefore only functions of the separation $r$ between the galaxies. 
The correlation of the convergences decreases sharply with the separation. This is due to the fact that, when the separation is small, the matter distributions along the two lines of sight (almost parallel for small $r$), which generate the lensing effects, are more likely to be correlated (if not even the same lenses when $r\approx0$). 
 In the same figure the auto-correlation of the convergence is also plotted for the configuration in which the two galaxies lie on the same line of sight with the middle point between them being at redshift $z_C=1$. 
 The reason why in this case the correlation of the convergences decreases much less rapidly with the separation is the following: the distance between the observer and any of the two sources is much bigger than the separation $r$ between the sources, therefore, the matter distribution causing the lensing effects is mostly that lying between the observer and the closer source.
As one can see from the figure the contributions from the spatial lapse and the velocity at the observer may have a non negligible effect on the correlation. While the spatial lapse at the observer cancels out in the full expression of the galaxy number density, the velocity at the observer does not and must be taken into account for both theoretical and numerical purposes.

\begin{figure}[h]
	\centering
		\includegraphics[width=\textwidth]{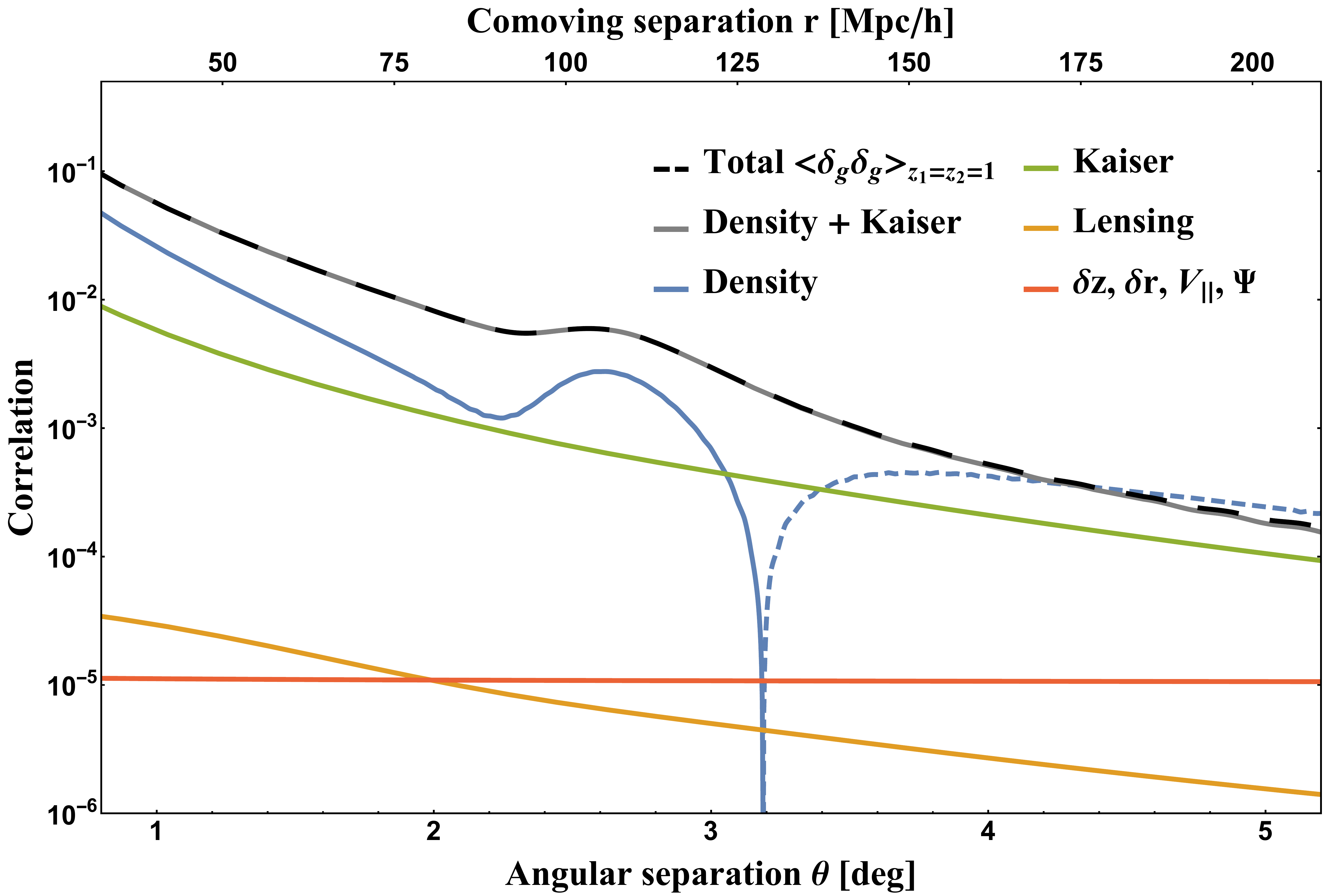}
		\caption{The full galaxy two-point correlation function $\langle \delta_g \delta_g \rangle$ for two galaxies at the same redshift $z_1=z_2=1$ is represented by the dashed black line. The gray line represents the standard correlation function that takes into account only the matter density contrast and the Kaiser effect. Other lines represent auto-correlations of various contributions to $\langle \delta_g \delta_g \rangle$: the matter density fluctuation in blue, the Kaiser effect in green, the gravitational lensing convergence in orange and the sum of all other effects in red. The latter is mostly influenced by the velocities effect. The dashed blue line represents negative values of the density correlation function. The value of the galaxy bias factor is set to $b=2$. The cut-off choices are $k_{\mathrm{IR}}=\mathcal H_o$ and $k_{\mathrm{UV}}=10$ $h$/Mpc.}
	\label{fig tot}
\end{figure} 
\begin{figure}[h]
	\centering
		\includegraphics[width=\textwidth]{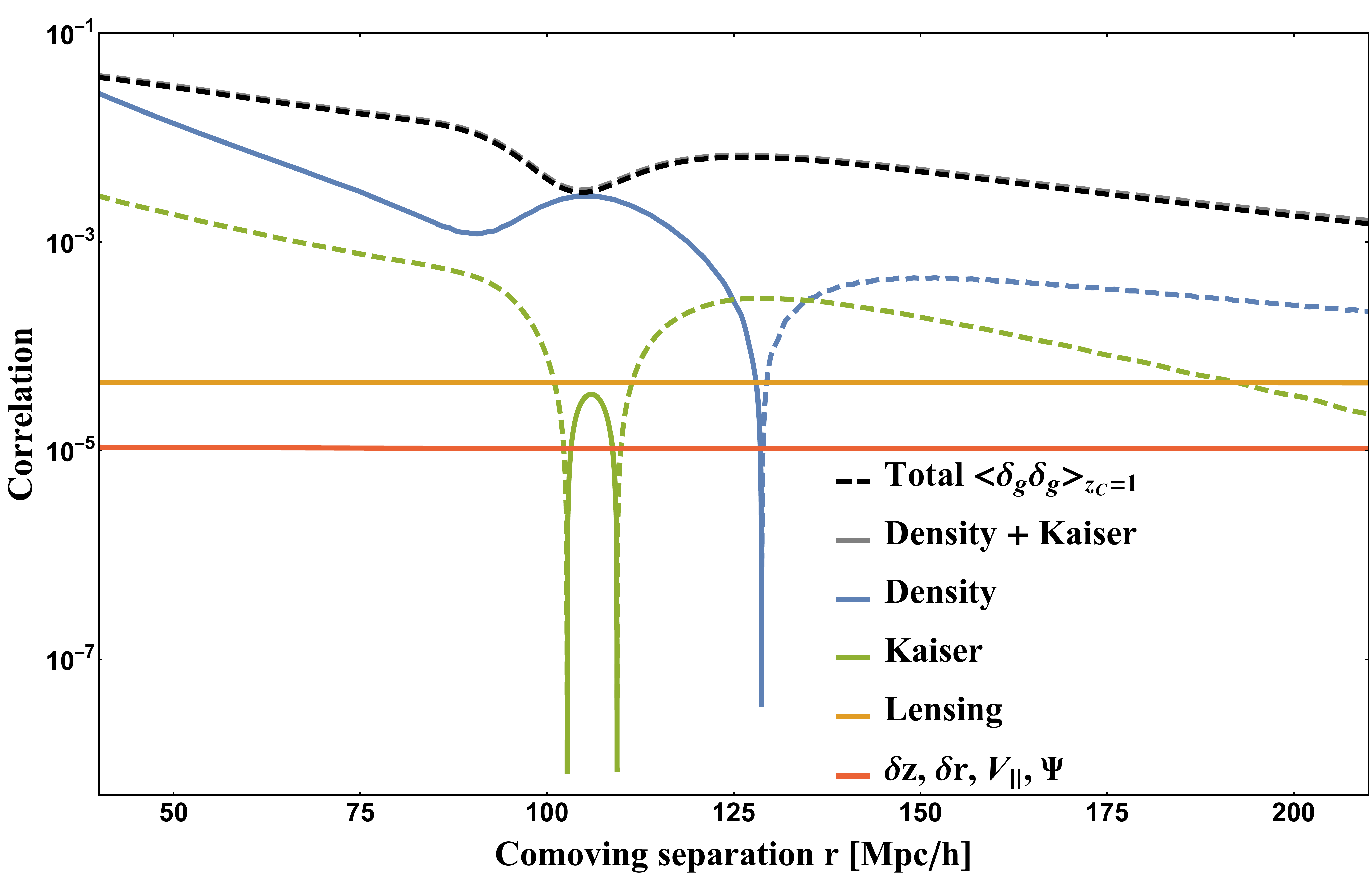}
		\caption{The full general relativistic two-point galaxy correlation function $\langle \delta_g \delta_g \rangle$ for two galaxies lying on the same line-of-sight $(\boldsymbol{\hat n}_1 = \boldsymbol{\hat n}_2 )$ with middle point at redshift $z_C=1$ is represented by the black line. The gray line represents the standard correlation function that takes into account only the matter density contrast and the Kaiser effect. Other lines represent auto-correlations of various contributions to $\langle \delta_g \delta_g \rangle$: the matter density fluctuation in blue, the Kaiser effect in green, the gravitational lensing convergence in orange and the sum of all other effects in red. The latter is mostly influenced by the velocities effect. The dashed lines represents negative values of the correlation functions. The value of the galaxy bias factor is set to $b=2$. The cut-off choices are $k_{\mathrm{IR}}=\mathcal H_o$ and $k_{\mathrm{UV}}=10$ $h$/Mpc.}
	\label{fig tot los}
\end{figure} 
\begin{figure}[h]
	\centering
		\includegraphics[scale=0.27]{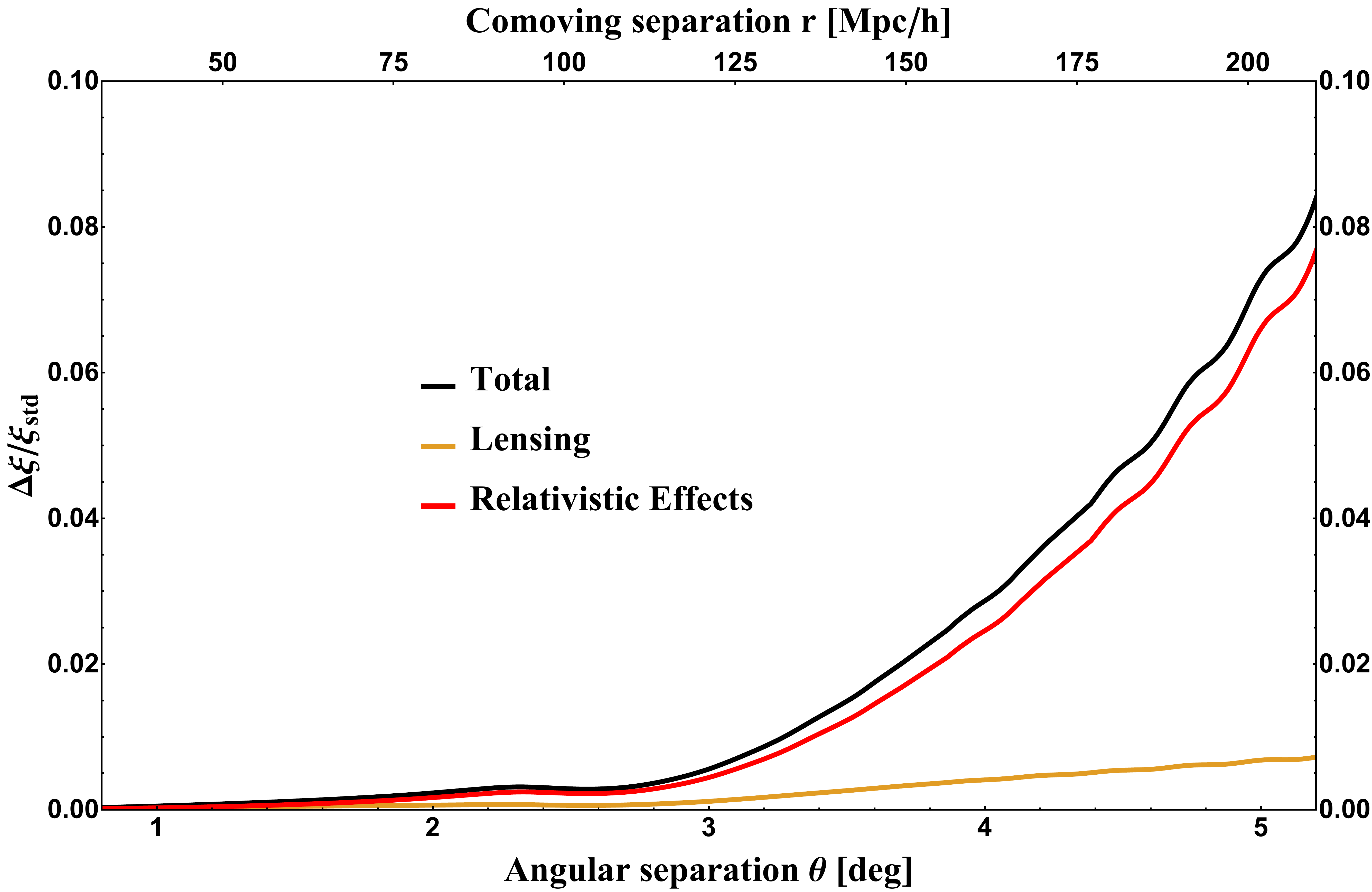}
		\caption{The black curve shows the relative difference between the full relativistic galaxy two-point correlation and the standard correlation of density and Kaiser effect only: $(\langle \delta_g \delta_g \rangle -\xi_{\text{std}})/\xi_{\text{std}}$, where $\xi_{\text{std}}\equiv \langle \delta_\text{std} \,\delta_\text{std}\rangle$ and $\delta_\text{std} = b\,\delta_m - \mathcal H_z^{ -1}\partial_{||}V_{||}$. Both $\langle \delta_g \delta_g \rangle$ and $\xi_{\text{std}}$ are computed by considering two galaxies at the same redshift $z_1=z_2=1$, so that the relative difference is only a function of the separation between the galaxies.
		The orange curve shows the contribution from lensing to the relative difference. Specifically, the orange line represents the relative difference between the correlation of lensing, density and Kaiser effect and the standard correlation: $(\xi_{\text{lensing}}-\xi_{\text{std}})/\xi_{\text{std}}$, where $\xi_{\text{lensing}}\equiv \langle (\delta_\text{std} - 2\, \mathcal K )(\delta_\text{std} - 2\, \mathcal K )\rangle$.  
The difference between the black and the orange curves, shown by the red curve, represents the
pure relativistic corrections to the standard theoretical predictions : $(\xi_{\text{rel}}-\xi_{\text{std}})/\xi_{\text{std}}$, where $\xi_{\text{rel}}=\langle (\delta_{\text{std}}+\delta_{\text{vel}}+\delta_{\text{pot}})(\delta_{\text{std}}+\delta_{\text{vel}}+\delta_{\text{pot}}) \rangle$ and $\delta_{\text{vel, pot}}$ are given in eq.~\eqref{var dg}. In this plot the galaxy bias is $b=2$ and the cut-off choices are $k_{\mathrm{IR}}=\mathcal H_o$ and $k_{\mathrm{UV}}=10$ $h$/Mpc. Note that $\xi_{\text{rel}}$ is independent of gauge choice.
}
	\label{fig diff}
\end{figure}

\subsubsection{Redshift-space distortions}\label{rsd}

Finally, in order to obtain the two-point correlation function of the galaxy number density fluctuation in eqs.~\eqref{delta g num} and \eqref{delta V num}, we have to consider the remaining term $H_z \frac{\partial}{\partial z} \delta r$ appearing in the volume distortion. By taking the derivative of the radial distortion with respect to the source redshift, this quantity can be written in terms of the redshift distortion $\delta z$ as
\begin{equation}
\label{eq tot kaiser}
	H_z \frac{\partial}{\partial z} \delta r = -\frac{1}{\mathcal H_z}\partial_{||}V_{||} - V_{||} + \frac{1}{\mathcal H_z} \Psi' + 2\,\Psi - \frac{\mathcal H_z'}{\mathcal H_z^2}\, \delta z \,.
\end{equation}
The first term represents the so-called redshift-space distortions (RSD), also referred to as the Kaiser effect. 
The auto-correlation function of this contribution to the galaxy clustering is given by
\begin{equation}\label{RSDRSD}
\begin{split}
	\frac{\langle \partial_{||}V_{||}(z_1,\boldsymbol{\hat n}_1) \partial_{||}V_{||}(z_2,\boldsymbol{\hat n}_2) \rangle}{\mathcal H_{z_1}\mathcal H_{z_2}} 
		= \frac{C^2 D_V(z_1)D_V(z_2)}{\mathcal H_{z_1}\mathcal H_{z_2}} \int_{k_\text{IR}}^{k_\text{UV}}\frac{dk}{2\pi^2} k^2 P_m(k) \bigg\lbrace   j_0''''(kr) \, \mu_1^2 \, \mu_2^2  
	\\
	+ \bigg[ \frac{j_0''(kr)}{(kr)^2}- \frac{j_0'(kr)}{(kr)^3} \bigg]\big(1 + 2\, \mu^2 - 3\, \mu_1^2 - 3\, \mu_2^2  -12\, \mu \, \mu_1 \, \mu_2 + 15 \, \mu_1^2 \, \mu_2^2 \big) 
	\\
		+ \frac{j_0'''(kr)}{kr} \big(\mu_1^2  + \mu_2^2 + 4 \,\mu \, \mu_1 \, \mu_2  - 6 \, \mu_1^2 \, \mu_2^2 \big)
	  \bigg\rbrace  \,,
\end{split} 
\end{equation}
where we have defined the angles $\mu \equiv \boldsymbol{\hat n_1} \cdot \boldsymbol{\hat n_2}$,\, $\mu_1 \equiv \boldsymbol{\hat n_1} \cdot \boldsymbol{\hat r}$,\, $\mu_2 \equiv \boldsymbol{\hat n_2} \cdot \boldsymbol{\hat r}$. 
Fig.~\ref{fig dens kaiser} shows the correlation as a function of the comoving separation $r$ between two galaxies at redshift $z_1=z_2=1$ and between two galaxies on the same line of sight $\boldsymbol{\hat n_1}=\boldsymbol{\hat n_2}$ with middle point at redshift $z_C=1$. In the configuration $z_1=z_2=1$ the correlation of the Kaiser effect has roughly the same amplitude of the matter densities correlation (if $b=1$ as in the figure). This is the reason why this is the only effect taken into account in the standard galaxy  correlation function, as the other effects are at least two orders of magnitude smaller. In the configuration $\boldsymbol{\hat n_1}=\boldsymbol{\hat n_2}$ and $z_C=1$ the correlation of the Kaiser effect is negative for almost all values of $r$, with a positive maximum at the scale of BAO. The reason why the BAO peak manifests only in the second configuration is that in this case the Kaiser effect of the two galaxies is related to same line of sight and therefore the correlation is sensitive to the clustering caused by the BAO, while in the first configuration the two lines of sight are different and arbitrary. 
Note that we use  ``the Kaiser effect'' to represent the contribution of the velocity gradient $-\partial_{||}V_{||}/\mathcal H_z$ only, rather than the sum of the velocity gradient and the density. 
The complete lack of a correlation between the lines of sight along which the Kaiser effect is evaluated removes the bump due to the BAO, so that the correlation simply decreases monotonically with the separation, independently from the clustering of matter.
Clearly, for $r=0$ the amplitude of the correlation is the same for the two configurations. However, the correlations are highly oscillatory for values of $r$ smaller than 35 Mpc/$h$ and, therefore, we only plot the functions starting from that separation value. 
The other terms in eq.~\eqref{eq tot kaiser} are much smaller than
the Kaiser effect, as one can see from fig.~\ref{fig dist}.

\subsubsection{Total two-point correlation function}
 
The total correlation function of the galaxy number density fluctuation $\delta_g$ for two galaxies at redshift $z_1=z_2=1$ is shown in fig.~\ref{fig tot}, while that for two galaxies lying on the same line-of-sight with middle point at redshift $z_C=1$ is shown in fig.~\ref{fig tot los}. Note that the pure relativistic contributions, represented by the red lines in both figures, are dominated by the velocity effects. These contributions are roughly the same in the configuration ($z_1=z_2$) of fig.~\ref{fig tot} and the configuration ($n_1=n_2$) of fig.~\ref{fig tot los}. From fig.~\ref{fig V} we see that the velocity correlation function in the two configurations are of the same order of magnitude, but in the configuration $n_1=n_2$ the velocity correlation at the source positions becomes negative at around 105 Mpc/$h$. However, the contribution of the velocity at the observer position, which is positive and greater than that at the source, makes the correlation in the two configurations being roughly the same.

From figs.~\ref{fig tot}$-$\ref{fig tot los} we readily recognize that
on most scales there exists little
difference between the full relativistic two-point correlation function $\langle \delta_{g} \delta_{g} \rangle$ and the standard correlation function $\xi_{\text{std}}=\langle \delta_{\text{std}} \, \delta_{\text{std}} \rangle$ that takes into account only the matter density and the Kaiser effect $(\delta_{\text{std}} \equiv b\,\delta_m - \mathcal H_z^{-1}\partial_{||}V_{||})$.
We further quantify this difference in fig.~\ref{fig diff}
for the configuration $z_1=z_2=1$.
Figure~\ref{fig diff} illustrates the fractional difference of the full 
relativistic description compared to the standard correlation function, 
and the orange curve shows the lensing contribution to this difference.
More importantly, fig.~\ref{fig diff} shows that
at large separation $r \approx 200$ Mpc/$h$ ($\theta \approx 5 \deg$)
the general relativistic effects cause corrections to the standard
 correlation function at several percent level, comparable or larger than the
lensing contribution. Such relativistic correction are mainly due to the velocity contribution $\delta_{\text{vel}}$ (see eq.~\eqref{var dg}), which is in turn dominated by the velocity at the observer position. The latter is often neglected in literature, leading to a systematic misinterpretation of the relativistic corrections. For separations smaller than the BAO scale the relativistic corrections, including lensing, are below the 1\% level, so that it can be legit to use the standard expression to analyze upcoming data, provided that the survey precision is not better than 1\%. For separations larger than 125 Mpc/h, instead, the lensing and the velocity contributions must be taken into account in the theoretical prediction, otherwise a systematic error of a few percents would affect the analysis. The potential contribution is small and can be neglected on all scales from the numerical point of view (see figs.~\ref{fig cut-off div} and \ref{fig dist}), but only once its theoretical importance is understood and under control. Indeed, the potential contribution is necessary for the gauge invariance of the expression and the consistency with the equivalence principle, also indispensable for the convergence of the correlation function in the infrared.

\subsection{Primordial gravitational wave contributions}
\label{tensor contribution}

In this section we investigate the various contributions to the two-point galaxy correlation function considering only tensor perturbations, corresponding to the primordial gravitational waves. In this case the expression of the observer galaxy number density fluctuation is derived in sec.~\ref{sec: tensor} as
\begin{equation}\label{dg tensor}
	\delta_g = (3-e_z)\,\delta z + 2\, \frac{\delta r}{\bar r_z} - 2\,\kappa  + H_z \frac{\partial}{\partial z} \delta r \,,
\end{equation}
where there is no tensor contribution to the matter density fluctuation and the relativistic distortions are given in terms of the projected tensor perturbations $C_{||} \equiv C_{ij}\hat n^i \hat n^j$ by
\begin{equation}\label{dzt drt}
\begin{split}
& \delta  z =   \int_0^{\bar r_z} d\bar r \,  {C_{||}}' \,, \qquad\qquad 
	\frac{\delta r}{\bar r_z} = - \frac{\delta z}{\bar r_z \mathcal H_z} -  \int_0^{\bar r_z} \frac{d\bar r}{\bar r_z}\, C_{||}  \,,
\\
		& \kappa =     \frac{5}{2} C_{||o} - C_{||}   - 3 \int_0^{\bar r_z}\frac{d\bar r}{ \bar r}
		    C_{||}   - \int_0^{\bar r_z}d\bar r\,   {C_{||}}' 
		  - \frac{1}{2} \int_0^{\bar r_z}d\bar r\, \bigg(\frac{\bar r_z - \bar r}{\bar r_z \bar r}\bigg)
		   \hat\nabla^2 C_{||}	\,,
		  \\
		&  H_z \frac{\partial}{\partial z} \delta r = - \frac{\mathcal H'_z}{\mathcal H^2_z} \delta z - \frac{1}{\mathcal H_z}{C_{||}}'   - C_{||}\,.
	\end{split}
\end{equation}
Note that the expression of the lensing convergence $\kappa$ in eq.~\eqref{kappa tens} has been manipulated by using the following relations:
\begin{equation}
\begin{split}
	& \hat\nabla_i =  \bar r(\delta^k_i - \hat n_i \hat n^k ) \partial_k = \bar r( \partial_i -  \hat n_i \hat n^k  \partial_k ) \,, \quad \hat\nabla^2 = \hat\nabla_i \hat\nabla^i = \bar r^2 \Delta - 2\,\bar r \, \hat n^k \partial_k - \bar r^2 \, \hat n^k \hat n^l \partial_k \partial_l \,,
\\
	& \hat n^k \partial_k = \partial_{\bar\eta} + \frac{d}{d\bar r} \,.
\end{split}
\end{equation}
In this way,  the tensor perturbations $C_{ij}$ appear through the contraction $C_{||}$ in all terms and in all expressions.

When all contributions in eq.~\eqref{dzt drt} are substituted into eq.~\eqref{dg tensor}, the expression of the galaxy number density can be reordered as
\begin{equation}\label{dg tensor total}
\begin{split}
\delta_g &= -5 C_{||o} + C_{||}   - \frac{1}{\mathcal H_z}{C_{||}}'  +6 \int_0^{\bar r_z}\frac{d\bar r}{ \bar r}
		    C_{||}     -  2  \int_0^{\bar r_z}\frac{d\bar r}{\bar r_z} C_{||}    
\\
& \quad   
	+ \bigg[(3-e_z) + 2  -  \frac{2}{\bar r_z \mathcal H_z}  - \frac{\mathcal H'_z}{\mathcal H^2_z}\bigg]  \int_0^{\bar r_z} d\bar r \,  {C_{||}}' 	  + \int_0^{\bar r_z}d\bar r\, \bigg(\frac{\bar r_z - \bar r}{\bar r_z \bar r}\bigg)
		   \hat\nabla^2 C_{||} \,,
\end{split}
\end{equation}
consistently with eqs.~(40)$-$(41) in \cite{DJFS LSS with tensors}. Note the presence of the observer term $C_{||o}$, due to the fact that we have set the initial conditions for integrating the geodesic equation by requiring that angular positions match the physical ones measured in the observer rest frame. In other words, such term represents the mismatch between the observer and the FRW coordinate systems, caused by tensor perturbations. As we have discussed in the previous section, considering observer terms in the case of scalar perturbations is important to guarantee the gauge invariance of the expressions, to ensure the convergence of the correlations in the infrared, and to obtain the correct amplitudes in the numerical evaluations. 
Despite the fact that there is no gauge ambiguity for tensor perturbations, considering the observer term $C_{||o}$ is essential for the consistency of the expressions with the equivalence principle. Indeed, as we have demonstrated in sec.~\ref{sec: tensor}, without the observer term the tensor contribution to the galaxy number density would contain the unphysical effects of uniform gravity from long-mode perturbations.
Furthermore, it has been already shown in \cite{DJFS LSS with tensors} that the observer term $C_{||o}$ is numerically important for the quadrupole of the observed galaxy number density. If such term is neglected, the tensor contribution to the quadrupole of the galaxy density cannot be estimated correctly.

In order to compute the two-point correlations, we first decompose the tensor perturbations into Fourier modes of two polarization states (labeled by $s=+,\times$) as in eq.~\eqref{tensor dec},
\begin{equation}\label{hij}
	C_{ij}(\eta,\boldsymbol{k}) = e_{ij}^+(\boldsymbol{\hat k}) \, C^+(\eta,\boldsymbol{k}) + e_{ij}^\times(\boldsymbol{\hat k}) \, C^\times(\eta,\boldsymbol{k}) \,,
\end{equation}
where the polarization tensors $e^s_{ij}(\boldsymbol{\hat k})$ are transverse, traceless and normalized through $e^s_{ij}e^{s'\,ij}=2\,\delta^{ss'}$. 
The power spectra of the two polarizations $C^+$ and $C^\times$ are 
\begin{equation}\label{polarizations}
	\langle C_s(\eta_1,\boldsymbol{k}_1) \, C_{s'}(\eta_2,\boldsymbol{k}_2) \rangle = (2\pi)^3 \delta_D(\boldsymbol{k}_1-\boldsymbol{k}_2)\delta_{ss'}\,\frac{1}{16}\,P_T(k_1,\eta_1,\eta_2)\,,
\end{equation}
where $P_T$ is the total tensor power spectrum $\propto \langle 2 C_{ij} \, 2C^{ij} \rangle$.
From eqs.~\eqref{hij} and \eqref{polarizations} the two-point correlation of tensor perturbations in terms of the power spectrum is given by
\begin{equation}
\langle C_{ij}(\eta_1,\boldsymbol{k}_1) C_{kl}(\eta_2,\boldsymbol{k}_2) \rangle = (2\pi)^3 \delta_D(\boldsymbol{k}_1- \boldsymbol{k}_2) \big[e_{ij}^+(\boldsymbol{\hat k}_1)e_{kl}^+(\boldsymbol{\hat k}_1) + e_{ij}^\times(\boldsymbol{\hat k}_1)e_{kl}^\times(\boldsymbol{\hat k}_1)\big]\frac{1}{16}P_T(k_1,\eta_1,\eta_2)\,.
\end{equation}
The tensor power spectrum can be further expressed in terms of the primordial one as
\begin{equation}\label{PT1}
	P_T(k,\eta_1,\eta_2)=T(k,\eta_1)\,T(k,\eta_2)\,P_{T0}(k)\,,
\end{equation}
where $T(k,\eta)$ is the tensor transfer function and the primordial power spectrum is given by an amplitude $A_T$ and an index $n_T$ as
\begin{equation}\label{PT2}
	P_{T0}(k) = \frac{2\pi^2}{k^{3}} \, A_T\, \bigg( \frac{k}{k_0} \bigg)^{n_T}.
\end{equation}
The amplitude can be obtained from that of the scalar modes as $A_T = r \, A_s$, where $r$ is the tensor-to-scalar ratio at the pivot scale $k_0$ and $A_s=2 \times 10^{-9}$. The index is also obtained from the tensor-to-scalar ratio as $n_T=-r/8$. Assuming $r = 0.2$ at $k_0 =0.003
$ $h/$Mpc, we have $A_T = 4 \times 10^{-10} $ and $n_T=-0.025$.
As we consider no anisotropic stress which sources gravitational waves, the tensor modes generated after inflation propagate freely. Thus, in the matter dominated epoch the transfer function is given by
\begin{equation}
	T(k,\eta) = 3 \frac{j_1(k\eta)}{k\eta} \,.
\end{equation}
This is still a valid approximation in the present epoch of accelerated expansion, and we will use it in the numerical calculations of the correlation functions. 

We now study the two-point correlation functions of the tensor contributions. We first write down the analytic expressions of the correlations of each relativistic distortion in eqs.~\eqref{dzt drt} for the general case. Then we study the correlations numerically as functions of the comoving separation between the two galaxies, considering only the configuration in which both galaxies are at redshift $z=1$. In this case, the correlation can be also expressed as a function of the angular separation $\theta$ between the two lines of sight.
However, the approach we have used to compute the correlation functions of the scalar contributions turns out to be complicated when applied to the tensor perturbations, as in this case the time and space dependence cannot be separated. Therefore, here we derive the correlation functions of the tensor contributions in terms of the angular power spectrum $C_l$
\begin{equation}\label{corr l}
\langle A(z_1,\bm{\hat n}_1) B(z_2,\bm{\hat n}_2) \rangle =  \xi_{AB}(z_1,z_2,\theta) = \frac{1}{4\pi}\sum_l (2l+1) C_l^{AB}(z_1,z_2) P_l(\cos\theta) \,,
\end{equation}
where $A$ and $B$ represent any of the relativistic corrections to the galaxy number density, such as $\delta z$, $\delta r/\bar r_z$, $\kappa$, $H_z \partial_z \delta r$,  and $P_l(x)$ are the Legendre polynomials. 

Note that each term in eq.~\eqref{dg tensor} or \eqref{dg tensor total} can be written as
\begin{equation}
\begin{split}
A(z,\bm{ \hat n}) &= \int_0^{\bar r_z} d\bar r\, W_A( z, \bar r) C_{||}(\bar \eta, \bar r \bm{ \hat n}) = \int_0^{\bar r_z} d\bar r\, W_A( z, \bar r) \int \frac{d^3k}{(2\pi)^3} e^{i \bar r \bm k \cdot \bm{ \hat n}}C_{||}(\bar \eta, \bm{ k})\,,
\\
&\equiv \int \frac{d^3k}{(2\pi)^3}A (z,\bm{\hat n},\bm{ k}) \,,
\end{split}
\end{equation}
where for $A=C_{||}$ then $W_A( z, \bar r)=\delta_D(\bar r - \bar r_z)$, for $A=\delta z$ then $W_A C_{||}= \partial_{\bar \eta} \, C_{||} \vert_{\bar \eta_o - \bar r}$ and so on.
First, we consider the contribution of a single plane-wave tensor perturbation, propagating along the $\bm{\hat z}$-direction ($\bm{\hat k}\equiv\bm{\hat z}$). Thus, the projection along the line of sight can be written as
\begin{equation}
	\hat n^i \hat n^j C_{ij}(\bar\eta,\boldsymbol{k}) =  \sin^2\theta [\cos2\phi \, C^+(\bar\eta,\boldsymbol{k}) + \sin2\phi \, C^\times(\bar\eta,\boldsymbol{k})]  = \sin^2\theta [e^{i 2 \phi} C_{+2} + e^{-i 2 \phi} C_{-2}]\,.
\end{equation}
Note that the helicity states are related to the polarization states as $C_{\pm 2} = \frac{1}{2} (C_+ \mp i C_\times)$, and their power spectra as $\langle C_{+2}C_{+2} \rangle = \langle C_{-2}C_{-2} \rangle = \frac{1}{2}\langle C_{+}C_{+} \rangle= \frac{1}{2}\langle C_{\times}C_{\times} \rangle$, while $\langle C_{+2}C_{-2} \rangle = 0$.
The contribution to $A(z,\bm{ \hat n})$ from this perturbation is given by
\begin{equation}
A (z,\bm{\hat n},\bm{ k}) = \int_0^{\bar r_z} d\bar r \, W_A( z, \bar r)\, e^{i k \bar r \mu} (1-\mu^2) [e^{i 2 \phi} C_{+2} (\bar \eta, \bm{ k}) + e^{-i 2 \phi} C_{-2} (\bar \eta, \bm{ k})]\,,
\end{equation}
where $\mu = \bm{\hat n} \cdot \bm{\hat k}$. The multipole coefficients of $A(z,\bm{ \hat n})$ are then
\begin{equation}
a^{A}_{lm} (z) = \int d^2 \hat n Y^*_{lm}(\bm{\hat n}) A (z,\bm{\hat n})  =\int \frac{d^3k}{(2\pi)^3} a^{A}_{lm}(z,\bm k) \,,
\end{equation}
where the multipole coefficient in Fourier space is
\begin{equation}
\begin{split}
a^{A}_{lm}(z,\bm k) &= \int d^2 \hat n Y^*_{lm}(\bm{\hat n}) A (z,\bm{\hat n},\bm{ k}) 
\\
&=\int_0^{\bar r_z} d\bar r \, W_A( z, \bar r)\,  \int d^2 \hat n Y^*_{lm}(\mu,\phi) e^{i k \bar r \mu} (1-\mu^2) [e^{i 2 \phi} C_{+2} (\bar \eta, \bm{ k}) + e^{-i 2 \phi} C_{-2} (\bar \eta, \bm{ k})] \,.
\end{split}
\end{equation}
By using the identity \cite{DJFS LSS with tensors}
\begin{equation}
\int d\Omega \, Y^*_{lm}(1-\mu^2)e^{\pm i 2\phi}e^{ix\mu} = - \sqrt{4\pi (2l+1)}\sqrt{\frac{(l+2)!}{(l-2)!}}i^l \frac{j_l(x)}{x^2}\delta_{m\pm 2}\,,
\end{equation}
the latter can be written as
\begin{equation}
a^{A}_{lm}(z,\bm k) = -i^l \sqrt{4\pi (2l+1)}\sqrt{\frac{(l+2)!}{(l-2)!}} \int_0^{\bar r_z} d\bar r\, W_A(\bar r,\bar r_z)\,  [C_{+2} (\bar \eta, \bm{ k})\delta_{m2} +  C_{-2} (\bar \eta, \bm{ k}) \delta_{m-2}]  \frac{j_l(k \bar r)}{(k \bar r)^2}\,.
\end{equation}

We now have all ingredients to compute the angular power spectrum $C_l$ and the two-point correlation functions  by using eq.~\eqref{corr l}. We have
\begin{equation}
\begin{split}
C_l^{AB}(z_1,z_2) &= \frac{1}{2l+1} \sum_m \mathrm{Re} \langle a^{A *}_{lm}(z_1) a^{B}_{lm} (z_2) \rangle \,,
\end{split}
\end{equation}
where the individual components with $A \equiv B$ are
\begin{equation}\label{cldz}
\begin{split}
C_l^{\delta z}(z_1,z_2) = \frac{1}{8\pi}  \frac{(l+2)!}{(l-2)!} \int dk \, k^2  P_{T0}(k) &  \int_0^{\bar r_{z_1}} d\bar r_1 \,  \frac{\partial}{\partial\bar \eta_1} T(k,\bar\eta_1)  \frac{j_l(k \bar r_1)}{(k \bar r_1)^2}
\\
 \times & \int_0^{\bar r_{z_2}} d\bar r_2 \,  \frac{\partial}{\partial\bar \eta_2} T(k,\bar\eta_2)  \frac{j_l(k \bar r_2)}{(k \bar r_2)^2}  \,,
\\
C_l^{\delta r}(z_1,z_2) = \frac{1}{8\pi}  \frac{(l+2)!}{(l-2)!} \int dk \, k^2  P_{T0}(k) &  \int_0^{\bar r_{z_1}} d\bar r_1 \, \bigg[- \frac{1}{\bar r_{z_1} \mathcal H_{z_1}} \frac{\partial}{\partial\bar \eta_1} - \frac{1}{\bar r_{z_1}} \bigg]T(k,\bar\eta_1)  \frac{j_l(k \bar r_1)}{(k \bar r_1)^2}
\\
 \times & \int_0^{\bar r_{z_2}} d\bar r_2 \,  \bigg[- \frac{1}{\bar r_{z_2} \mathcal H_{z_2}} \frac{\partial}{\partial\bar \eta_2} - \frac{1}{\bar r_{z_2}} \bigg] T(k,\bar\eta_2)   \frac{j_l(k \bar r_2)}{(k \bar r_2)^2} \,,
\\
C_l^{\kappa}(z_1,z_2) = \frac{1}{8\pi}  \frac{(l+2)!}{(l-2)!} \int dk \, k^2  P_{T0}(k)  &
\\
\times  \int_0^{\bar r_{z_1}} d\bar r_1 \, \bigg[ \frac{5}{2}\delta_D(\bar r_1) - \delta_D(\bar r_1 - \bar r_{z_1} &)  - \frac{3}{\bar r_1} -  \frac{\partial}{\partial \bar\eta_1} + \frac{ l (l+1)}{2} \frac{\bar r_{z_1} - \bar r_1}{\bar r_{z_1} \bar r_1} \bigg]T(k,\bar\eta_1)  \frac{j_l(k \bar r_1)}{(k \bar r_1)^2}
\\
 \times  \int_0^{\bar r_{z_2}} d\bar r_2 \,  \bigg[ \frac{5}{2}\delta_D(\bar r_2) - \delta_D(\bar r_2  - \bar r_{z_2} &)  - \frac{3}{\bar r_2} -  \frac{\partial}{\partial \bar\eta_2} + \frac{ l (l+1)}{2} \frac{\bar r_{z_2} - \bar r_2}{\bar r_{z_2} \bar r_2} \bigg] T(k,\bar\eta_2)   \frac{j_l(k \bar r_2)}{(k \bar r_2)^2} \,,
\\
C_l^{\partial_z \delta r}(z_1,z_2) = \frac{1}{8\pi}  \frac{(l+2)!}{(l-2)!} \int dk \, k^2  P_{T0}(k) & 
\\
\times \int_0^{\bar r_{z_1}} d\bar r_1 \, \bigg[ - \frac{\mathcal H_{z_1}'}{\mathcal H_{z_1}^2}\frac{\partial}{\partial \bar\eta_1}  - & \frac{1}{\mathcal H_{z_1}}\delta_D(\bar r_1 - \bar r_{z_1})\frac{\partial}{\partial \bar\eta_1}  - \delta_D(\bar r_1 - \bar r_{z_1}) \bigg]T(k,\bar\eta_1)  \frac{j_l(k \bar r_1)}{(k \bar r_1)^2}
\\
 \times  \int_0^{\bar r_{z_2}} d\bar r_2 \,  \bigg[ - \frac{\mathcal H_{z_2}'}{\mathcal H_{z_2}^2}\frac{\partial}{\partial \bar\eta_2}  - & \frac{1}{\mathcal H_{z_2}}\delta_D(\bar r_2 - \bar r_{z_2})\frac{\partial}{\partial \bar\eta_2}  - \delta_D(\bar r_2 - \bar r_{z_2}) \bigg] T(k,\bar\eta_2)   \frac{j_l(k \bar r_2)}{(k \bar r_2)^2} \,.
\end{split}
\end{equation}
Note that the time derivatives of the transfer functions are evaluated at $\bar \eta= \bar \eta_o - \bar r$ and we have used the relation $a_{lm}[\hat\nabla^2 A] = -l(l+1)a_{lm}^A$ to obtain $C_l^{\kappa}$. The angular power spectra for $A \neq B$ are obtained analogously.
The total tensor contribution to the angular power spectrum is then given by
\begin{equation}
\begin{split}
C_l^{\text{tot}}(z_1,z_2) =& \frac{1}{8\pi}  \frac{(l+2)!}{(l-2)!} \int dk \, k^2  P_{T0}(k) 
\\
\times   \int_0^{\bar r_{z_1}} d\bar r_1 \,& \bigg[(3-e_z) W_{\delta z}(\bar r_1) + 2 W_{\delta r} (z_1,\bar r_1) - 2 W_{\kappa}(z_1,\bar r_1,l) + W_{\partial_z \delta r}(z_1,\bar r_1) \bigg] T(k,\bar\eta_1)  \frac{j_l(k \bar r_1)}{(k \bar r_1)^2}
\\
 \times  \int_0^{\bar r_{z_2}} d\bar r_2 \, &\bigg[(3-e_z) W_{\delta z}(\bar r_2) + 2 W_{\delta r} (z_2,\bar r_2) - 2 W_{\kappa}(z_2,\bar r_2,l) + W_{\partial_z \delta r}(z_1,\bar r_2) \bigg] T(k,\bar\eta_2)   \frac{j_l(k \bar r_2)}{(k \bar r_2)^2} \,,
\end{split}
\end{equation}
where $W_{\delta z}$, $W_{\delta r}$, $W_{\kappa}$ and $W_{\partial_z \delta r}$ are read off eq.~\eqref{cldz}.
Note that $C_l=0$ for $l=\lbrace 0,1 \rbrace$, as for tensor perturbations the only scalar that can be constructed out of $C_{ij}$ is the contraction $\hat n^i \hat n^j C_{ij}$, whose multipole expansion starts from the quadrupole. 

\begin{figure}
	\centering
		\includegraphics[width=\textwidth]{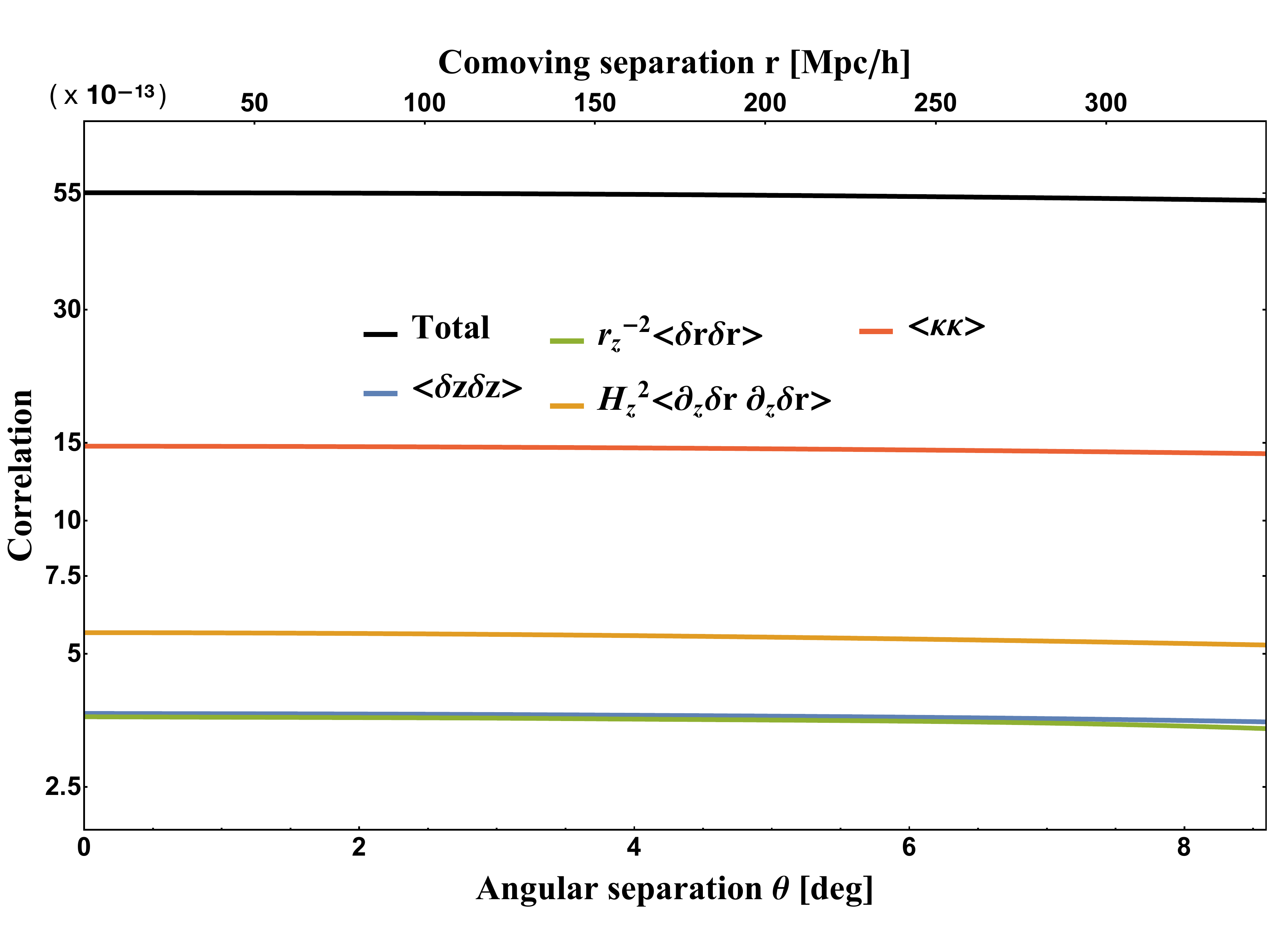}
		\caption{Two-point correlations of the relativistic contributions to the galaxy number density due to primordial gravitational waves. The correlations are functions of the separation between two galaxies at the same redshift $z_1=z_2=1$.}
           	\label{fig tensor}
\end{figure}

In fig.~\ref{fig tensor} we summarize our numerical results, obtained by considering two galaxies at the same redshift $z_1=z_2=1$. The two-point correlations are therefore functions of the comoving separation $r$ between the two spatial positions or the angular separation $\theta$ between the two lines of sight. 
The plot shows that the total tensor contribution to the two-point galaxy correlation function is of order $10^{-12}$, and it varies very little with the separation. Among the relativistic corrections the lensing convergence is the most important, being four times larger than the contributions from the redshift and the radial distortions.
The amplitude of the correlation functions is expected to be small, as primordial gravitational waves decay fast once they enter the horizon. We find, indeed, that the effect of gravitational waves is suppressed by eight or more orders of magnitude  with respect to the scalar contributions.

\section{Summary and Discussion}
\label{discussion}

In this work we have studied the two-point galaxy correlation function, both theoretically and numerically, providing the complete general relativistic predictions at linear order that are essential to interpret its measurements. 
Many groups (e.g. \cite{Jeong,Yoo:2012,Hui:2007cu,LoVerde:2007ke,Hui:2007tm,Raccanelli:2013dza,Raccanelli:2013gja,Raccanelli:2015vla,Vitto,Tansella:2018sld}) have already presented the relativistic galaxy two-point correlation function, considering different cosmologies and exploring broad redshift intervals with various configurations of the galaxy pairs. However, this work addresses and resolves theoretical issues concerning the expression of the galaxy number density and its two-point correlation function. 
Following the lead of \cite{DJFS LSS with tensors,cosmic rulers,SG divergence,
SG correlation,Jeong},
we have shown that the boundary terms evaluated at the observer position are necessary for the gauge-invariance of the expression, for its consistency with the equivalence principle and for the convergence of the correlation function in the infrared regime. 

The galaxy number density is an observable, measured by counting the number of galaxies in the survey volume. As such, its theoretical expression has to be independent from the gauge condition chosen for the computations. We have derived the theoretical expression, starting from a general metric representation with scalar and tensor perturbations, without imposing any gauge choice. In this way, we could explicitly verify the gauge-invariance of our expression and check its validity. It is important to stress that the gauge-invariant expression is obtained by deriving the observed galaxy number density 
in terms of physical quantities, namely the observed redshift and the angular position on the sky. These quantities are measured in the observer rest frame, 
which requires the frame change from the FRW coordinates and results in
perturbation contributions at the observer position. These boundary terms
at the observer position cannot be set zero as it is often done in literature. The perturbations evaluated at the observer position are, indeed, necessary for the gauge-invariance of the expression. 

Furthermore, this complete gauge-invariant expression including the boundary
terms is needed for the compatibility with the equivalence principle. 
In sec.~\ref{EP} we have demonstrated that our expression satisfies the equivalence principle by showing explicitly that it is unaffected by the uniform gravitational potential and the uniform acceleration generated by long-mode 
scalar or tensor 
perturbations, i.e. perturbations with wavelength much larger than the distance between the observer and the source, representing the scale of the system. 
When considering the 
two-point galaxy correlation function, the infrared divergences generated by the monopole of the gravitational potential at the source and integrated along the line of sight are cancelled by the divergent contributions at the observer,
providing a finite result
 (see figures \ref{fig cut-off div} and \ref{fig variances}).
If the perturbations at the observer position are set zero by hand, 
the divergent contributions are not balanced in the two-point correlation
function, 
and one is forced to impose an {\it arbitrary} infrared cut-off when computing correlations.

One might argue that the perturbation contributions at the observer position are constants and, therefore, taking an ensemble average to correlate them is conceptually incorrect. 
Furthermore, the real observer only takes spatial average over the sources. Since the ergodic theorem provides a correspondence between the spatial averaging and the theoretical ensemble averaging, one might argue that the ensemble average should not be taken over the perturbations at the observer, as there is no corresponding spatial average \cite{Tansella:2018sld}. Indeed, we do not have access to measurements taken from different observers in the universe.
However, the perturbations at the observer are random fields evaluated at a point, exactly as perturbations at the source.
Furthermore, the distinction of the perturbation contributions at the observer position and the rest is a gauge-dependent interpretation.
In the conformal Newtonian gauge we adopted for the computation there exist
perturbation contributions at the observer position. However, 
in the comoving gauge, for instance, there are no perturbation contributions at the observer position, but those at the observer position in the conformal Newtonian gauge are instated as the perturbations at the source position.
As a consequence, one cannot treat the observer position differently from any other point, when the ensemble average is taken. This approach is the only way that leads to a theoretically consistent result in any gauge conditions.

Adopting the conformal Newtonian gauge, we have performed numerical studies
of the individual relativistic contributions to the galaxy two-point
correlation function. The contributions to the observed galaxy number
density are divided into the redshift distortion~$\delta z$, the radial 
distortion~$\delta r$, the gravitational lensing convergence~$\cal K$, 
and the Kaiser effect (or the redshift-space distortion). 
In such decomposition, each contribution is gauge dependent and some of them are IR diverging in the correlation function. However, since we have shown that the sum is gauge invariant and its correlation converges in the infrared, we have imposed an IR cutoff for the purpose of illustration.
With this, we have computed the relativistic contributions to the galaxy two-point correlation
function, considering two configurations of galaxy pairs: the one in which the two galaxies are at the same redshift $z=1$ (transverse), and the one in which the two galaxies are along the same line-of-sight with middle point at fixed 
redshift $z=1$ (parallel).

Our numerical results reproduce the standard two-point correlation function, which accounts for the density fluctuation and the RSD, in complete agreement with the literature (see fig.~\ref{fig dens kaiser}). It is interesting to note that the auto-correlation of the RSD exhibit the BAO feature in the parallel configuration, but not in the transverse one, as for the latter there is no correlation between the two lines of sight. The standard expression is used to analyze data from current surveys, as the precision of such measurements does not require higher theoretical accuracy. However, for future surveys the sub-percent level of accuracy is demanded by the increasing precision of data, and the theoretical expression must include all the relativistic effects.
Our numerical results show that the gravitational lensing convergence represents the most important relativistic effect after the RSD, for small angular separations ($\theta < 2$ deg) in the configuration where both galaxies are at redshift $z=1$ (see fig.~\ref{fig tot}) and for any comoving separation in the other configuration (see fig.~\ref{fig tot los}).
The correlations of other relativistic effects are dominated by the effect of peculiar velocities (see fig.~\ref{fig dist}).
In particular, the contribution from the velocity at the observer is the most important (see fig.~\ref{fig V}), but it is often ignored in the literature. 

A detailed analysis of the correlation function was performed in \cite{Vitto}.
The bottom left panel of fig.~3 in \cite{Vitto} provides the fractional errors
due to the individual relativistic contributions to the correlation function
in the same format as our fig.~\ref{fig diff}. Compared to the standard calculation
$\xi_{\text{std}}$ in fig.~\ref{fig diff}, the relativistic contribution is 6\% at $r=200$ Mpc/h,
largely due to the velocity contribution. However, we find a factor 10
difference in fig.~3 in \cite{Vitto}, where the velocity contribution (blue)
is 0.6\% at the same separation. Apart from the factor two difference in
galaxy bias, the cosmological parameters adopted in \cite{Vitto} and our analysis
are fairly identical. However, we note that the calculation of the correlation
function in \cite{Vitto} neglects all the contributions at the observer position,
and the velocity contribution among those at the observer position is factor
10 larger than the source velocity contribution at $r=200$ Mpc/h
shown in fig.~\ref{fig V}. We believe that the factor 10 difference in the fractional
errors can be attributed to the missing velocity contribution
at the observer position. The gravitational potential contribution (green)
in \cite{Vitto} appears larger than the velocity contribution (blue), whereas
the potential contributions in our calculation are typically three orders
of magnitude smaller than the velocity contributions. The more recent study in \cite{Tansella:2018sld} reported the same results as in \cite{Vitto}, so that all points of the above comparison with our study applies also to \cite{Tansella:2018sld}.

As mentioned above, one cannot neglect the boundary terms at the observer in the expression of the observable galaxy number density. While the other perturbations at the observer (the time-lapse and the gravitational potential) are important mostly because they eliminate the unobservable and divergent monopole from the correlation, as their contribution has a very small amplitude compared with the density and RSD, the velocity at the observer contributes to the dipole of the correlation and has a non-negligible effect. Since the correlation of velocities at the observer is almost constant, it is particularly important for large separations, where the correlation of other contributions decreases. 
Also the spatial shift at the observer $\delta r_o$ would contribute to the dipole, but it cancels exactly in the theoretical expression. It is important to consider it, however, to correctly predict the correlations of radial distortions, for which it represents the leading contribution, and lensing convergences (fig.~\ref{fig dist}).
We emphasize again that these individual quantities are gauge-dependent, such that the separation of the correlation function into these terms is not unique and ignoring any of these terms would lead to an inconsistent result in a different gauge choice. 
Indeed, the observable two-point correlation function is only the total one, and the theoretical (gauge-invariant) sum of various (gauge-dependent) contributions has to match it in any gauge.
For instance, one can choose the comoving gauge, in which
the individual relativistic corrections would contribute differently, but the sum is the same as in the conformal Newtonian gauge. Note, however, that the individual relativistic corrections in the comoving gauge would also diverge differently than in the conformal Newtonian gauge, so that the gauge invariance of the expression is not a sufficient condition. Indeed, note that not all gauge-invariant expressions describe actual physical observables and that an expression may be gauge-invariant but not compatible with the equivalence principle.

By computing the total correlation, we have also shown that ignoring relativistic effects on top of the density fluctuation and the RSD would lead to a relative error that can reach the 8\% for two galaxies at redshift $z=1$ separated by 5 $\deg$ at the observer (see fig.~\ref{fig diff}). This means that one should use the relativistic expression to interpret future data from upcoming surveys on such large scales. The terms involving the gravitational potential (including Sachs-Wolf and Shapiro time-delay effects) have a negligible contribution to the amplitude, at least 5 orders of magnitude smaller then the (leading) density contribution, and can be ignored. This holds also when the luminosity distance is concerned, as in this case the potential contribution is much smaller than the (leading) velocity one.  However, since the terms involving the gravitational potential are cut-off dependent in the infrared, one can neglect them numerically only when their role is understood and theoretically under control. Our work serves also this purpose, providing the correct description of all relativistic effects in the galaxy two-point correlation function.

Finally, we have calculated the correlations of individual relativistic corrections due to the primordial gravitational waves and their total contribution to the two-point galaxy correlation function. 
Since the galaxy number density is affected by gravitational waves via redshift and volume distortions, the two-point galaxy correlation function can be used as a probe for the primordial gravitational waves predicted by inflation.
Unfortunately, tensor modes decay inside the horizon, so that their effect is only important at large scales and high redshifts.
Consequently, the tensor contribution to the two-point galaxy correlation function is very small, in particular compared to the scalar contribution that grows in time.
In our numerical study we have considered the configuration where both galaxies are at redshift $z=1$ and the correlation is a function of the angular separation (see fig.~\ref{fig tensor}). As expected, our results show that with a tensor-to-scalar ratio of 0.2 the tensor contribution is of order $10^{-12}$, which is about eight or more orders of magnitude smaller than the scalar contribution, making it difficult to detect the primordial gravitational waves with galaxy clustering. 

We have provided theoretical and numerical studies of the full relativistic two-point galaxy correlation function. A deep understanding of all theoretical subtleties in the relativistic description of galaxy clustering is essential to interpret the numerous upcoming surveys. 
Indeed, only the correct theoretical prediction can lead us to the full realization of the cosmological potential of galaxy clustering enabled by precision measurements in future galaxy surveys.

\acknowledgments

We thank Ermis Mitsou, Giuseppe Fanizza, Nastassia Grimm, and Vittorio Tansella for useful discussions. 
 We acknowledge support by the Swiss National Science Foundation, and J.Y. is further supported by a Consolidator Grant of the European Research Council (ERC-2015-CoG grant 680886).

\appendix

\section{$\Lambda$CDM solutions for scalar perturbations}
\label{app solutions}

At linear order, all Fourier modes grow at the same rate and the time dependence of the scalar perturbations in the conformal Newtonian gauge can be expressed in terms of the growth function $D$ of the linear density fluctuation $\delta(a,\boldsymbol{x})=D(a)\delta_+(\boldsymbol{x})$ and the curvature perturbation $\zeta(\boldsymbol{x})$ in the comoving gauge.
From the conservation of energy and momentum in a $\Lambda$CDM universe, one derives the evolution equation for the linear growth function $D$ as
\begin{equation}
	\frac{d^2 D}{da^2} + (2-\Omega_m)\frac{3}{2a}\frac{d D}{da}-\frac{3}{2a^2} D = 0 \,.
\end{equation}
The analytic solution is well-known:
\begin{equation}\label{D1}
	D(a)= a \, _2 F_1\bigg[\frac{1}{3},1,\frac{11}{6},-\frac{a^3}{\Omega_m}(1-\Omega_m)\bigg]\,,
\end{equation}
where $_2 F_1$ is the hypergeometric function and $\Omega_m=\Omega_m(a)$ is the matter density parameter.

Using the Einstein equations in the comoving gauge $(\gamma=v=0\,,\,\,\varphi \equiv \zeta)$, the perturbations can be expressed in terms of the spatial configuration $\delta_+(\boldsymbol{x})$ of the density contrast or the curvature perturbation $\zeta(\boldsymbol{x})$ as \cite{Yoo:2016tcz}
\begin{equation}
\begin{split}
\zeta(\boldsymbol x) &=  C\, \Delta^{-1} \, \delta_+(\boldsymbol x)\,,  
	\\
\beta(a, \boldsymbol x) &=  \frac{C}{\mathcal H \, \Sigma} \, \Delta^{-1} \, \delta_+(\boldsymbol x) = \frac{1}{\mathcal H \Sigma}\, \zeta(\boldsymbol x) \equiv D_\beta (a) \zeta (\boldsymbol x) \,,	
\end{split}
\end{equation}
where we defined the time-dependent functions
\begin{equation}
 D_\beta \equiv \frac{1}{\mathcal H\, \Sigma}\,, \quad\qquad \Sigma \equiv 1+ \frac{3}{2} \frac{\Omega_m}{f}\,, \quad\qquad f \equiv \frac{d\,\ln D}{d\,\ln a}\,.
\end{equation}
Since $\zeta$ is time-independent, $C$~is a constant
\begin{equation}
	C \equiv  - f \, D \, \mathcal H^2 \, \Sigma ~,\qquad\qquad
D(a)\propto{1\over{\cal H}^2f\Sigma}~,
\end{equation}
and it becomes $C=-\frac{5}{2}\mathcal H_o^2 \Omega_m$ in the Einstein-de
Sitter universe.

The perturbation solutions in the conformal Newtonian gauge with no anisotropic pressure $(\beta=\gamma=0\,,\,\, \alpha = - \varphi \equiv \Psi )$ are obtained by transforming the solution in the comoving gauge. Given the gauge-transformations in sec.~\ref{metric} one obtains $T=\beta$ and $L=0$. Therefore, the perturbation variables in the conformal Newtonian gauge are related to the comoving gauge variables as 
\begin{equation}
\begin{split}
	&\Psi = \frac{1}{a}(a \, \beta)', \qquad\quad 
	\Psi = - \mathcal H \beta - \zeta \,, 
	\qquad\quad 
	v =  -\beta\,, 
	\\
	&\delta\eta_o = - v_{o}\,, \qquad\quad
	\delta x^{i}_{o}=-\int_0^{\bar \eta_o}d\bar \eta\, {v}^{\,,i}\,.
\end{split}
\end{equation}
These can be further written in terms of the curvature perturbation as
\begin{equation}
\begin{split}
	&\Psi(\eta,\boldsymbol{x}) = D_\Psi(\eta)\zeta(\boldsymbol{x})\,, \qquad\qquad v(\eta,\boldsymbol{x}) = -D_V(\eta)\zeta(\boldsymbol{x})\,, 
	\\
	& \delta\eta_o =  D_{Vo}\,{\zeta}_o\,, \qquad\qquad\qquad\qquad \delta x^{i}_{o}= \big({\zeta}^{,i}\big)_o \int_0^{\bar \eta_o}d\bar \eta\, D_V \,,
\end{split}
\end{equation}
where $D_\Psi=\mathcal H D_\beta -1 $ and $D_V=D_\beta$. By combining the above equations, we derive the relations 
\begin{equation}
\begin{split}
	& D_\Psi = - \mathcal H D_V - D_V' \,, \qquad D_\Psi = - \frac{1}{2}(D_V'+1)\,, \qquad \int_0^{\bar r_z}d\bar r\, D_\Psi = \frac{1}{2}(D_V - D_{Vo} - \bar r_z)\,,
	\\
	&  D_V' + 2 \, \mathcal H D_V - 1 = 0   \,, \qquad  D_\Psi' = \frac{\mathcal H'}{\mathcal H}(D_\Psi+1)-2\,\mathcal H (D_\Psi +1) + \mathcal H \,.
\end{split}	
\end{equation}

\section{Derivations of velocity and RSD correlation functions}
\label{app derivations}

The analytical expression for the two-point correlation function of the velocities at the sources in eq.~\eqref{v corr} is obtained by using the relations $V_{||}(\eta,\boldsymbol x)=D_V(\eta) \partial_{||} \zeta(\boldsymbol x)$ and $\zeta(\boldsymbol x) =  C\, \Delta^{-1} \, \delta_+(\boldsymbol x)$, derived in appendix \ref{app solutions}. We have
\begin{equation}
\begin{split}
	&\langle V_{||} (z_1, \hat n_1) V_{||}(z_2, \hat n_2) \rangle = D_V(\eta_1) D_V(\eta_2) \langle \hat n^i_1 \partial_i\zeta(\boldsymbol{x}_1) \, \hat n^j_2 \partial_j\zeta(\boldsymbol{x}_2) \rangle
		\\
	&= - C^2 D_V(\eta_1) D_V(\eta_2) \int \frac{d^3 k}{(2\pi)^3}  e^{i \boldsymbol{k} \cdot   \boldsymbol r}  (i \boldsymbol{\hat n}_1 \cdot \boldsymbol{k})(i \boldsymbol{\hat n}_2 \cdot \boldsymbol{k})\frac{P_m(k)}{k^4} 
	\\
	&= - C^2 D_V(\eta_1) D_V(\eta_2)\hat n_1^i \hat n_2^j  \int \frac{d^3 k}{(2\pi)^3}  \frac{\partial}{\partial r_i} \frac{\partial}{\partial r_j} e^{i \boldsymbol{k} \cdot   \boldsymbol r}   \frac{P_m(k)}{k^4} 
		\\
	&= - C^2 D_V(\eta_1) D_V(\eta_2)\hat n_1^i \hat n_2^j  \int \frac{d^3 k}{(2\pi)^3}  \frac{\partial}{\partial r_i}\bigg( k \hat r_j\frac{\partial}{\partial (kr)} e^{i \boldsymbol{k} \cdot   \boldsymbol r} \bigg)  \frac{P_m(k)}{k^4} 
			\\
	&= - C^2 D_V(\eta_1) D_V(\eta_2)\hat n_1^i \hat n_2^j  \int \frac{d k}{2\pi^2}  \bigg[ \frac{\delta_{ij}+\hat r_i \hat r_j}{kr}\frac{\partial}{\partial(kr)} j_0(kr) +  \hat r_i \hat r_j \frac{\partial^2}{\partial(kr)^2} j_0(kr) \bigg]  P_m(k)
	\\
	&= \bigg( \frac{C}{\mathcal H_o} \bigg)^2 D_V(z_1) D_V(z_2) \big\lbrace \hat{\mathcal P}_{||} \xi_{||}(r) + \hat{\mathcal P}_{\perp} \xi_{\perp}(r) \big\rbrace  \,,
\end{split}
\end{equation}
where we defined $\hat{\mathcal P}_{||} \equiv \hat n^i_1 \hat n^j_2\, \hat r_i \hat r_j$ and $\hat{\mathcal P}_{\perp} \equiv \hat n^i_1 \hat n^j_2\, (\delta_{ij}- \hat r_i \hat r_j)$ to decompose the velocity correlation function into the parallel and perpendicular components with respect to the separation $\boldsymbol{r}=\boldsymbol x_1 - \boldsymbol x_2$: 
\begin{equation}
	\xi_{||}(r) \equiv -\mathcal H_o^2 \int_{k_\text{IR}}^{k_\text{UV}}\frac{dk}{2\pi^2} P_m(k) \frac{j_0' (kr)}{kr}\,, \qquad\qquad \xi_{\perp}(r) \equiv -\mathcal H_o^2 \int_{k_\text{IR}}^{k_\text{UV}}\frac{dk}{2\pi^2} P_m(k) j_0'' (kr) \,,
\end{equation}
with $j_0'(x)=\partial_x j_0(x)$ and $j_0''(x)=\partial_x^2 j_0(x)$.

The analytical expression for the two-point correlation function of the  in eq.~\eqref{RSDRSD} is obtained by using the same relations as above. We have
\begin{equation}
\begin{split}
	&\frac{\langle \partial_{||}V_{||}(z_1,\boldsymbol{\hat n}_1) \partial_{||}V_{||}(z_2,\boldsymbol{\hat n}_2) \rangle}{\mathcal H_{z_1}\mathcal H_{z_2}} 
	= \frac{C^2 D_V(z_1)D_V(z_2)}{\mathcal H_{z_1}\mathcal H_{z_2}}\hat n^i_1 \hat n^j_1 \hat n^k_2 \hat n^l_2 \int \frac{d^3 k}{(2\pi)^3} (ik_i)(ik_j)(ik_k) (ik_l) e^{i \boldsymbol k \cdot  \boldsymbol r} \frac{P_m(k)}{k^4}
				\\
	&= \frac{C^2 D_V(z_1)D_V(z_2)}{\mathcal H_{z_1}\mathcal H_{z_2}}\hat n^i_1 \hat n^j_1 \hat n^k_2 \hat n^l_2 \int \frac{d k}{2\pi^2} \bigg[ k \hat r_i \frac{\partial}{\partial (kr)}  \bigg(k \hat r_j\frac{\partial}{\partial (kr)} \bigg(k \hat r_k \frac{\partial}{\partial (kr)}\bigg(k \hat r_l\frac{\partial}{\partial (kr)} j_0(kr) \bigg)\bigg)\bigg) \bigg]\frac{P_m(k)}{k^2}
	\\
	&= \frac{C^2 D_V(z_1)D_V(z_2)}{\mathcal H_{z_1}\mathcal H_{z_2}} \int_{k_\text{IR}}^{k_\text{UV}}\frac{dk}{2\pi^2} k^2 P_m(k) \bigg\lbrace   j_0''''(kr) \, \mu_1^2 \, \mu_2^2  + \frac{j_0'''(kr)}{kr} \big(\mu_1^2  + \mu_2^2 + 4 \,\mu \, \mu_1 \, \mu_2  - 6 \, \mu_1^2 \, \mu_2^2 \big)
	\\
	&\quad + \bigg[ \frac{j_0''(kr)}{(kr)^2}- \frac{j_0'(kr)}{(kr)^3} \bigg]\big(1 + 2\, \mu^2 - 3\, \mu_1^2 - 3\, \mu_2^2  -12\, \mu \, \mu_1 \, \mu_2 + 15 \, \mu_1^2 \, \mu_2^2 \big) 
	  \bigg\rbrace  \,,
\end{split} 
\end{equation}
where we have defined the angles $\mu \equiv \boldsymbol{\hat n_1} \cdot \boldsymbol{\hat n_2}$,\, $\mu_1 \equiv \boldsymbol{\hat n_1} \cdot \boldsymbol{\hat r}$,\, $\mu_2 \equiv \boldsymbol{\hat n_2} \cdot \boldsymbol{\hat r}$.

\end{document}